\documentclass[preprint,authoryear,12pt]{elsarticle}

\usepackage{graphicx}
\usepackage{amssymb}

\journal{International Journal of Solids and Structures}

\begin{document}

\begin{frontmatter}

\title{Green's formula and singularity at a triple contact line. Example of finite-displacement solution}

\author{Juan Olives}

\ead{olives@cinam.univ-mrs.fr}

\address{CINaM-CNRS Aix-Marseille Universit\'e, Campus de Luminy, case 913, 13288 Marseille cedex 9, France}

\begin{abstract}
The various equations at the surfaces and triple contact lines of a deformable body are obtained from a variational condition, by applying Green's formula in the whole space and on the Riemannian surfaces. The surface equations are similar to the Cauchy's equations for the volume, but involve a special definition of the `divergence' (tensorial product of the covariant derivatives on the surface and the whole space). The normal component of the divergence equation generalizes the Laplace's equation for a fluid--fluid interface. Assuming that Green's formula remains valid at the contact line (despite the singularity), two equations are obtained at this line. The first one expresses that the fluid--fluid surface tension is equilibrated by the two surface stresses (and not by the volume stresses of the body) and suggests a finite displacement at this line (contrary to the infinite-displacement solution of classical elasticity, in which the surface properties are not taken into account). The second equation represents a strong modification of Young's capillary equation. The validity of Green's formula and the existence of a finite-displacement solution are justified with an explicit example of finite-displacement solution in the simple case of a half-space elastic solid bounded by a plane. The solution satisfies the contact line equations and its elastic energy is finite (whereas it is infinite for the classical elastic solution). The strain tensor components generally have different limits when approaching the contact line under different directions. Although Green's formula cannot be directly applied, because the stress tensor components do not belong to the Sobolev space $H^1({\rm V})$, it is shown that this formula remains valid. As a consequence, there is no contribution of the volume stresses at the contact line. The validity of Green's formula plays a central role in the theory.
\end{abstract}

\begin{keyword}
Surface and contact line equations\sep Elasticity\sep Green's formula\sep Singularities
\end{keyword}

\end{frontmatter}

\section{Introduction}

Surface properties of deformable bodies have been continually studied since the early work of \cite{Gibbs:1878} until recent mechanical or thermodynamic studies, e.g.~\cite{Gurtin-etal:1998}, \cite{Simha-Bhattacharya:2000}, \cite{Rusanov:2005}, \cite{Steinmann:2008} and \cite{Olives:2010a}. They have many applications, e.g. in adhesion, coating and nanosciences (since small and thin objects are deformable and have a high surface/volume ratio). A previous paper \citep{Olives:2010a} was devoted to the physical basis of the theory: application of the equilibrium criterion of Gibbs; introduction of the new concept of `ideal transformation', i.e., the homogeneous extrapolation of the deformation, in the interface film, up to the dividing surface; determination of the thermodynamic variables of state of a surface; definition of the surface stress tensor; surface and line equations. Moreover, for an elastic solid, it is known that classical elasticity predicts a singularity with an infinite displacement (and an infinite elastic energy) at a solid--fluid--fluid triple contact line, owing to the fluid--fluid surface tension which is a force concentrated on this line \citep{Shanahan-deGennes:1986,Shanahan:1986}. Although some authors tried to overcome this problem, by introducing some fluid--fluid interface thickness \citep{Lester:1961,Rusanov:1975}, some cut-off radius near the contact line \citep{Shanahan-deGennes:1986,Shanahan:1986} or some new elastic force at this line \citep{Madasu-Cairncross:2004}, this situation makes very difficult to write any equilibrium equation at the contact line.

The present paper concerns the mathematical foundation of the theory. A sketch of the proof of the surface and contact line equations is presented (no proof was given in the previous physical paper \cite{Olives:2010a}), which shows (i) the importance of the validity of Green's formula at the contact line (despite the singularity) and (ii) owing to the surface properties, the probable existence of a finite-displacement solution (consequence of the line equations, based themselves on the assumption of the validity of Green's formula). These two points are justified with an explicit example of finite-displacement solution, in the simple case of a half-space solid, bounded by a plane, and subjected to a normal force concentrated on a straight line of its surface. This solution also shows that the elastic energy is finite and that Green's formula remains valid at the contact line.
 
\section{Surface and contact line equations} \label{secsurfaceline}

\begin{figure}[t]
\begin{center}
\includegraphics[height=4.5cm]{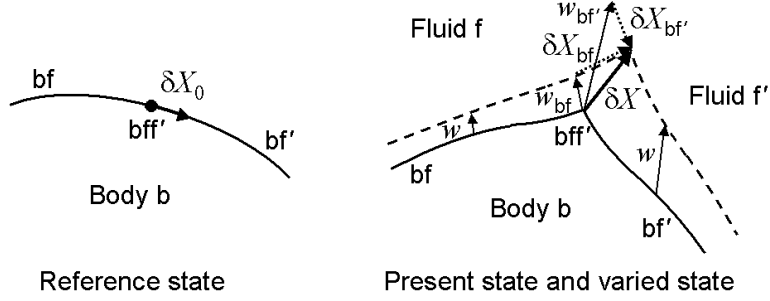}
\end{center}
\caption{Displacement $\delta X_0$ of the contact line $\rm bff'$ with respect to the body, in the reference state, and displacement $\delta X$ of this line in the space, between the present state and its varied state, due to both the displacement of the line with respect to the body and the displacement of the material points of the body (see text).} \label{deltaX}
\end{figure}

For a general deformable body $\rm b$ in contact with various immiscible fluids $\rm f$, $\rm f'$,... (with no mass exchange between the body and the fluids), the mechanical equilibrium condition relative to the body, including its body--fluid surfaces ($\rm bf$, $\rm bf'$,...) and its body--fluid--fluid triple contact lines ($\rm bff'$,...), may be written as
\begin{eqnarray}
&&\int_{\rm b} \pi : \delta e\,dv_0 
- \int_{\rm b} \rho\,\bar g \cdot \delta x\,dv \nonumber\\
&&- \sum_{\rm bf} \int_{\rm bf} p\,n \cdot \delta x\,da
- \sum_{\rm bf} \int_{\rm bf} \rho_{\rm s}\,\bar g \cdot \delta x\,da
+ \sum_{\rm bf} \int_{\rm bf} \pi_{\rm s} : \delta e_{\rm s}\,da_0 \nonumber\\
&&- \sum_{\rm bff'} \int_{\rm bff'} \gamma_{\rm ff'}\,\nu_{\rm ff'}\cdot \delta X\,dl 
+ \sum_{\rm bff'} \int_{\rm bff'} (\gamma_{0,{\rm bf}} - \gamma_{0,{\rm bf'}}) 
\,\delta X_0\,dl_0 = 0 \label{bodyvariational1}
\end{eqnarray}
(: means double contraction; see \cite{Olives:2010a} for the physical basis of the theory), in which $\delta$ is an arbitrary variation such that, on the closed surface $\Sigma$ which bounds the system, the points of the body and the points of the body--fluid--fluid lines remain fixed. In this expression, $\pi$ is the Piola--Kirchhoff stress tensor (i.e., the Lagrangian form, relative to a reference state of the body) at equilibrium, $e$ the Green--Lagrange strain tensor (also relative to this reference state), $dv$ and $dv_0$ are respectively the volume measures in the present state and in the reference state, $\rho$ is the mass per unit volume, $\bar g$ the (constant) gravity vector field, $\delta x$ the displacement of a point of the body, $p$ the fluid pressure, $n$ the unit vector normal to the $\rm bf$ surface, oriented from $\rm f$ to $\rm b$, $da$ and $da_0$ are respectively the area measures in the present state and in the reference state, $\rho_{\rm s}$ is the mass per unit area (excess on the dividing surface $\rm S_{bf}$ defined by the condition: no excess of mass of the constituent of the body), $\pi_{\rm s}$ the (Lagrangian) surface stress tensor at equilibrium, defined in \cite{Olives:2010a}, $e_{\rm s}$ the (Lagrangian) surface strain tensor, defined in \ref{secEulLag}, $\gamma_{\rm ff'}$ the fluid-fluid surface tension, $\nu_{\rm ff'}$ the unit vector normal to the $\rm bff'$ line, tangent to the $\rm ff'$ surface, and oriented from the line to the interior of $\rm ff'$, $\delta X$ the (vector) displacement of the $\rm bff'$ line, perpendicular to the line (in the present state), $dl$ and $dl_0$ are respectively the length measures in the present state and in the reference state, $\gamma_{0}$ is the surface grand potential (excess on the dividing surface), per unit area in the reference state, and $\delta X_0$ the (scalar) displacement of the $\rm bff'$ line, measured in the reference state, perpendicular to that line in the reference state, and positively considered from $\rm bf$ to $\rm bf'$ (see also Fig.~\ref{deltaX}, below).

This variational equilibrium condition leads to various equations at the surfaces and the triple contact lines of the body. Since the preceding paper \citep{Olives:2010a} was devoted to the physical aspects of the theory, these equations were only written without proof. In this section, a sketch of this proof is presented, which shows the importance of the validity of Green's formula to obtain the contact line equations (despite the line singularity). These equations then suggest the existence of a finite-displacement solution.

In order to only have quantities or variables (such as points, forces, etc.) which refer to the present equilibrium state in these equations, we first need to transform all the Lagrangian terms in (\ref{bodyvariational1}) (i.e., those related to the `undeformed' reference state) into Eulerian forms (i.e., related to the deformed present state). It is well known that the Eulerian forms of the above (volume) stress and strain tensors, $\pi$ and $\delta e$, are the Cauchy stress tensor $\sigma$ and the infinitesimal strain tensor $\delta \varepsilon$ defined below in the next paragraph (see e.g.~\cite{Mandel:1966}, tome I, annexe II). Note that $e$ measures the strain between the `undeformed' reference state and the deformed present state, so that its components $e_{ij}$ may have arbitrary values, since large strains may occur in highly deformable bodies (even when subjected to capillary forces or surface stresses). Note also that $\delta \varepsilon$, which measures the infinitesimal strain between the present state and its varied state (i.e., after the variation $\delta$), is not the variation of some strain tensor, but it is related to the variation $\delta e$ of the Lagrangian tensor $e$ by
\begin{eqnarray*}
\delta e = \Phi_0^* \cdot \delta \varepsilon \cdot \Phi_0,
\end{eqnarray*}
where $\Phi_0$ is the deformation gradient between the reference state and the present state and $\Phi_0^*$ its adjoint. Classically, the work of deformation of a volume element (first term of (\ref{bodyvariational1})) may be written in the Eulerian form
\begin{eqnarray}
\pi : \delta e\,dv_0 = \sigma : \delta \varepsilon\,dv \label{workdefvol}
\end{eqnarray}
(see \cite{Mandel:1966}, ibid.), i.e., with arbitrary Cartesian coordinates in the three-dimensional space $\rm E$
\begin{eqnarray*}
\pi^{ij}\,\delta e_{ij}\,dv_0 = \sigma^{ij}\,\delta \varepsilon_{ij}\,dv,
\end{eqnarray*}
where Latin indices $i$, $j$, $k$,... belong to $\{1,2,3\}$ and summation is performed over repeated indices. In a similar way, these concepts are extended to the surfaces in \ref{secEulLag}, where the Lagrangian surface strain tensor $e_{\rm s}$, the Eulerian infinitesimal surface strain tensor $\delta \varepsilon_{\rm s}$ and the Eulerian surface stress tensor $\sigma_{\rm s}$ are defined. The work of deformation of a surface element (fifth term of (\ref{bodyvariational1})) may then be written in the Eulerian form (\ref{workdefsurf}).

Let us first consider the simple case of a bounding surface $\Sigma$ which only encloses one fluid $\rm f$ and the body $\rm b$. The equilibrium condition (\ref{bodyvariational1}) may then be written as
\begin{eqnarray*}
&&\int_{\rm V} \sigma : \delta \varepsilon\,dv 
- \int_{\rm V} \rho\,\bar g \cdot w\,dv \\
&&- \int_{\rm S} p\,n \cdot w\,da
- \int_{\rm S} \rho_{\rm s}\,\bar g \cdot w\,da
+ \int_{\rm S} \sigma_{\rm s} : \delta \varepsilon_{\rm s}\,da = 0,
\end{eqnarray*}
where $w = \delta x$, $\rm V$ is the bounded open set of $\rm E$ occupied by the part of the body enclosed in $\Sigma$, and $\rm S$ the bounded part of $\rm S_{bf}$ enclosed in $\Sigma$. Since $\delta \varepsilon = \frac{1}{2} (({\rm D}w)^* +\,{\rm D}w)$ and
\begin{eqnarray*}
\int_{\rm V} {\rm tr}(\sigma \cdot \delta \varepsilon)\,dv
&=& \int_{\rm V} {\rm tr}(\frac{\sigma^* + \sigma}{2}\cdot {\rm D}w)\,dv\\
&=& \int_{\rm V} {\rm tr}(\sigma^* \cdot {\rm D}w)\,dv
+ \int_{\rm V} {\rm tr}(\frac{\sigma - \sigma^*}{2}\cdot {\rm D}w)\,dv,
\end{eqnarray*}
by application of Green's formula (with $w = 0$ on $\Sigma$)
\begin{eqnarray}
\int_{\rm V} {\rm tr}(\sigma^* \cdot {\rm D}w)\,dv
&=& - \int_{\rm V} {\rm div}(\sigma^*) \cdot w\,dv
- \int_{\rm S} (\sigma^* \cdot w) \cdot n\,da \nonumber\\
&=& - \int_{\rm V} {\rm div}(\sigma^*) \cdot w\,dv
- \int_{\rm S} (\sigma \cdot n) \cdot w\,da \label{Greenvolume}
\end{eqnarray}
(if the components of $\sigma$ and $w$ belong to $C^1(\bar {\rm V})$; e.g.~\cite{Allaire:2007}, Sec.~3.2.1), this leads to the classical Cauchy's equations for the body
\begin{eqnarray}
&&{\rm div}\,\bar\sigma + \rho\,\bar g = 0 \label{bequilibrium1}\\
&&\sigma^* = \sigma, \label{bequilibrium2} 
\end{eqnarray}
where ${\rm div}\,\bar\sigma$ is the vector associated to the linear form ${\rm div}(\sigma^*)$, and the remaining condition for the surface
\begin{eqnarray}
- \int_{\rm S} (\sigma \cdot n) \cdot w\,da
- \int_{\rm S} p\,n \cdot w\,da
- \int_{\rm S} \rho_{\rm s}\,\bar g \cdot w\,da
+ \int_{\rm S} \sigma_{\rm s} : \delta \varepsilon_{\rm s}\,da = 0,\label{Surfvar1}
\end{eqnarray}
for any variation such that $w = 0$ on the closed curve $\Gamma = \rm S_{bf} \cap \Sigma$ which bounds $\rm S$. Note that, if volume moments $\bar M\,dv$ were present, the new term $- \int_{\rm V} \frac{1}{2} {\rm tr}(M^* \cdot {\rm D}w)\,dv$\footnote{$M$ is the endomorphism defined by $(M \cdot x)\cdot y = [\bar M,x,y]$, for any vectors $x$ and $y \in \rm E$; it satisfies $M^* = - M$.} would appear in the equilibrium condition, and (\ref{bequilibrium2}) would become $\sigma - \sigma^* + M = 0$.

By applying Green's formula on the Riemannian manifold $\rm S$ \citep{Courrege:1966} to the last term of the equilibrium condition, we then obtain
\begin{eqnarray}
&&- \int_{\rm S} (\sigma \cdot n) \cdot w\,da
- \int_{\rm S} p\,n \cdot w\,da
- \int_{\rm S} \rho_{\rm s}\,\bar g \cdot w\,da \nonumber\\
&&- \int_{\rm S} {\rm div}({\widetilde{\sigma_{\rm s}}}^*) \cdot w\,da
+ \int_{\rm S} {\rm tr}(\frac{\sigma_{\rm s} - \sigma_{\rm s}^*}{2} \cdot \iota^* \cdot \psi)\,da = 0 \label{Surfvar2}
\end{eqnarray}
(see notations in \ref{secEulLag}), where $\widetilde{\sigma_{\rm s}} = \iota \cdot \sigma_{\rm s}$ and the special divergence of ${\widetilde{\sigma_{\rm s}}}^*$ is based on the tensorial product of the covariant derivative on $\rm T(S_{bf})$ and the usual derivative on $\rm S_{bf} \times E^*$ (see \ref{secSurfeq}). This leads to the following equations for the surface
\begin{eqnarray}
&&{\rm div}\,\bar{\sigma_{\rm s}} + \rho_{\rm s}\,\bar g + \sigma \cdot n + p\,n
= 0 \label{bfequilibrium1}\\
&&\sigma_{\rm s}^* = \sigma_{\rm s}, \label{bfequilibrium2} 
\end{eqnarray}
where ${\rm div}\,\bar{\sigma_{\rm s}}$ is the vector associated to the linear form ${\rm div}({\widetilde{\sigma_{\rm s}}}^*)$, i.e., according to (\ref{divergencecomp}),
\begin{eqnarray}
&&\partial_\beta (\sigma_{\rm s}^{\alpha\beta}\,\partial_\alpha x^i) +
\Gamma_{\beta\gamma}^\beta\,\sigma_{\rm s}^{\alpha\gamma}\,\partial_\alpha x^i +
\rho_{\rm s}\,\bar g^i + \sigma^{ij}\,n_j + p\,n^i = 0 \nonumber\\
&&\sigma_{\rm s}^{\beta\alpha} = \sigma_{\rm s}^{\alpha\beta} \label{bfequilibrium1comp}
\end{eqnarray}
(in the two last equations, $\sigma_{\rm s}$ and $\sigma$ as contravariant tensors, by raising the covariant index to the second place). Note the similarity of these equations with the classical Cauchy's ones (\ref{bequilibrium1}) and (\ref{bequilibrium2}) for the volume. Note also that (\ref{bfequilibrium2}) might be different if surface moments were present (as for the volume stress: see the comment after (\ref{Surfvar1})). The above Eq. (\ref{bfequilibrium1}) has a tangential component
\begin{eqnarray}
{\rm div}\,\sigma_{\rm s} + \rho_{\rm s}\,{\bar g}_t + (\sigma \cdot n)_t
= 0 \label{bfequilibrium1tangent}
\end{eqnarray}
(${\rm div}\,\sigma_{\rm s}$ being the usual surface divergence; the subscript $t$ indicates the vector component tangent to $\rm S_{bf}$) and a normal component
\begin{eqnarray}
l_n : \sigma_{\rm s} + \rho_{\rm s}\,{\bar g}_n + \sigma_{nn} + p
= 0, \label{bfequilibrium1normal}
\end{eqnarray}
where $l_n = l \cdot n$ ($l_{n,\alpha\beta} = l_{\alpha\beta}^i\,n_i$; $l$ is the second vectorial fundamental form on $\rm S_{bf}$), ${\bar g}_n = \bar g \cdot n$ and $\sigma_{nn} = (\sigma \cdot n) \cdot n$ (see \ref{secSurfeq}). At any point $x \in \rm S_{bf}$, the eigenvalues of $l_n$ (as endomorphism) are the principal curvatures, $\frac{1}{R_1}$ and $\frac{1}{R_2}$, of $\rm S_{bf}$ (\cite{Dieudonne:1971}, (20.14.2); a curvature being positive when its centre is on the side of $n$). Note that, if $\sigma_{\rm s}$ is isotropic, i.e., $\sigma_{\rm s} = \hat{\sigma_{\rm s}}\,I$ (eigenvalue $\hat{\sigma_{\rm s}}$ and $I$ the identity), then ${\rm div}\,\sigma_{\rm s} = {\rm grad}\,\hat{\sigma_{\rm s}}$ and $l_n : \sigma_{\rm s} = \hat{\sigma_{\rm s}}\,{\rm tr}(l_n) = \hat{\sigma_{\rm s}} (\frac{1}{R_1} + \frac{1}{R_2})$. In particular, if the deformable body $\rm b$ is a fluid, the application of the general thermodynamic equations (26), (27) and (29) of \cite{Olives:2010a} and (19) of \cite{Olives:2010b}, and their comparison with the classical fluid--fluid equations (see (1), (12) in \cite{Olives:2010a}), leads to $\sigma_{\rm s} = \gamma\,I$, thus $\hat{\sigma_{\rm s}} = \gamma$. This is also a consequence of (12) of \cite{Olives:2010b}, since $\gamma$ (for a fluid--fluid surface) does not depend on the surface strain $\varepsilon_{\rm s}$. In this particular case, the above Eq. (\ref{bfequilibrium1normal}) leads to the classical Laplace's equation for a fluid--fluid interface and (\ref{bfequilibrium1tangent}) to the classical hydrostatic equilibrium for the surface tension, $d\gamma = \rho_{\rm s}\,g\,dz$ \citep{Gibbs:1878} ($g$ is the norm of $\bar g$ and $z$ the height). The above Eqs. (\ref{bfequilibrium1tangent}) and (\ref{bfequilibrium1normal}) are then a generalization of these classical equations.

Similar surface equations were obtained for elastic solids from a balance of momentum or equilibrium of forces \citep{Moeckel:1975,Gurtin-Murdoch:1975,Simha-Bhattacharya:2000,Javili-Steinmann:2010}, a virtual power method \citep{Daher-Maugin:1986}, a thermodynamic approach \citep{Alexander-Johnson:1985,Leo-Sekerka:1989}, or an energy minimization \citep{Gurtin-etal:1998,Steinmann:2008}. In these works, the existence of a surface stress tensor was often assumed, deduced from a given surface traction field \citep{Gurtin-Murdoch:1975}, or defined for elastic solids from a given set of thermodynamic or mechanical variables of state of the surface \citep{Alexander-Johnson:1985,Leo-Sekerka:1989,Gurtin-etal:1998,Steinmann:2008}. Note that our thermodynamic method \citep{Olives:2010a}, valid for any deformable body (such as a viscoelastic solid, a viscous fluid or any other one) and based on the general equilibrium criterion of Gibbs, leads to the determination of the `local' thermodynamic variables of state of the surface, the definition of the surface stress tensor and the above equations. Note also that the divergence term in (\ref{bfequilibrium1}) is here defined as a true divergence with respect to a special covariant derivative, i.e. the tensorial product of the covariant derivatives on the surface and the whole space (in previous works, this term was only defined by means of its scalar product with a constant vector).

\begin{figure}[t]
\begin{center}
\includegraphics[width=5cm]{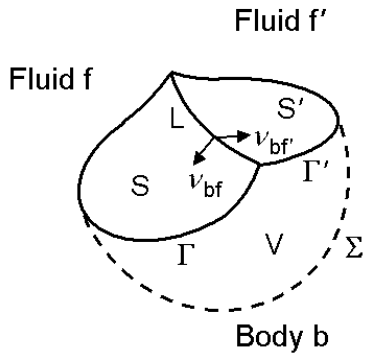}
\end{center}
\caption{The bounding surface $\Sigma$ encloses the parts $\rm V$, $\rm S$, $\rm S'$ and $\rm L$ of, respectively, the body $\rm b$, the surface $\rm bf$, the surface $\rm bf'$ and the triple contact line $\rm bff'$ (the part of $\Sigma$ in contact with the fluids and the surface $\rm ff'$ are not represented).} \label{figVSS'L}
\end{figure}

Let us now apply the equilibrium condition (\ref{bodyvariational1}) with a bounding surface $\Sigma$ which encloses two fluids, $\rm f$ and $\rm f'$, and the body $\rm b$, in contact. $\rm V$ denotes the bounded open set of $\rm E$ occupied by the part of the body enclosed in $\Sigma$, $\rm S$ the bounded part of $\rm S_{bf}$ enclosed in $\Sigma$, $\rm S'$ the bounded part of $\rm S_{bf'}$ enclosed in $\Sigma$, and $\rm L$ the part of the $\rm bff'$ triple contact line enclosed in $\Sigma$ (Fig.~\ref{figVSS'L}).
We follow the same method as above, but Green's formula (\ref{Greenvolume}) cannot be directly applied on $\rm V$ ($\rm S$ being here replaced with $\rm S \cup S'$), owing to the singularity at the contact line. If $\rm b$ is a deformable solid subjected to a force concentrated on a line of its surface (here, the fluid--fluid surface tension $\gamma_{\rm ff'}$ applied on the contact line), then classical elasticity predicts a singularity with an infinite displacement at this line, together with an infinite value of the elastic energy \citep{Shanahan-deGennes:1986,Shanahan:1986}. Nevertheless, we shall see, in this paper, that the introduction of the surface properties leads to a solution with a finite displacement at the contact line and a finite value of the elastic energy. In the example of finite-displacement solution presented in the next section, the singularity at the contact line involves components of $\sigma$ which do not belong to $H^1(\rm V)$. Although Green's formula cannot be directly applied in this case (we would need that components of both $\sigma$ and $w$ belong to $H^1(\rm V)$; e.g.~\cite{Allaire:2007}, Sec.~4.3.3), we show in Sec.~\ref{secvalidGreen} that this formula remains valid. We may thus assume that Green's formula is valid and, following the above method, the remaining condition for the surfaces and the line becomes (with the help of (\ref{bequilibrium1}) and (\ref{bequilibrium2}))
\begin{eqnarray*}
&&- \int_{\rm S \cup S'} (\sigma \cdot n) \cdot w\,da
- \int_{\rm S \cup S'} p\,n \cdot w\,da
- \int_{\rm S \cup S'} \rho_{\rm s}\,\bar g \cdot w\,da \\
&&+ \int_{\rm S \cup S'} \sigma_{\rm s} : \delta \varepsilon_{\rm s}\,da 
- \int_{\rm L} \gamma_{\rm ff'}\,\nu_{\rm ff'}\cdot \delta X\,dl 
+ \int_{\rm L_0} (\gamma_{0,{\rm bf}} - \gamma_{0,{\rm bf'}})\,\delta X_0\,dl_0 = 0
\end{eqnarray*}
($\rm L_0$ is the position of $\rm L$ in the reference state), for any variation such that $w = 0$ on the curves $\Gamma = \rm S_{bf} \cap \Sigma$ and $\Gamma' = \rm S_{bf'} \cap \Sigma$ which bound $\rm S \cup S'$, and the two points of $\rm L$ which belong to $\Sigma$ remain fixed. Note that there is no singularity at $\rm L_0$ in the reference state of the body (which is, e.g., a state of the body before its contact with the fluids $\rm f$ and $\rm f'$). The application of (\ref{sigmas-sigmasstar}) and Green's formula (\ref{Greensurface}) to the two terms $\int_{\rm S} \sigma_{\rm s} : \delta \varepsilon_{\rm s}\,da + \int_{\rm S'} \sigma_{\rm s} : \delta \varepsilon_{\rm s}\,da$ leads to the two new terms $- \int_{\rm L} (\sigma_{\rm s,bf} \cdot \nu_{\rm bf}) \cdot w_{\rm bf}\,dl - \int_{\rm L} (\sigma_{\rm s,bf'} \cdot \nu_{\rm bf'}) \cdot w_{\rm bf'}\,dl$ (subscripts $\rm bf$ and $\rm bf'$ respectively denote the sides of $\rm S = bf$ and $\rm S' = bf'$; thus, $\nu_{\rm bf}$ is the unit vector tangent to $\rm S_{bf}$, normal to $\rm L$ and directed to the inside of $\rm S_{bf}$; similarly, for $\nu_{\rm bf'}$ with respect to $\rm S_{bf'}$; see Fig.~\ref{figVSS'L}; $w = 0$ on $\Gamma$ and $\Gamma'$, but not on the $\rm bf$ and $\rm bf'$ sides of $\rm L$) and then, with the help of (\ref{bfequilibrium1}) and (\ref{bfequilibrium2}) on $\rm S$ and $\rm S'$, to the remaining line condition
\begin{eqnarray*}
&&- \int_{\rm L} (\sigma_{\rm s,bf} \cdot \nu_{\rm bf}) \cdot w_{\rm bf}\,dl 
- \int_{\rm L} (\sigma_{\rm s,bf'} \cdot \nu_{\rm bf'}) \cdot w_{\rm bf'}\,dl \\
&&- \int_{\rm L} \gamma_{\rm ff'}\,\nu_{\rm ff'}\cdot \delta X\,dl 
+ \int_{\rm L_0} (\gamma_{0,{\rm bf}} - \gamma_{0,{\rm bf'}})\,\delta X_0\,dl_0 = 0,
\end{eqnarray*}
for any variation such that the two points of $\rm L$ which belong to $\Sigma$ remain fixed (both in the space and with respect to the body $\rm b$). Since the displacement $\delta X$ of the contact line in the space is due to both the displacement of the corresponding material points of the body ($w_{\rm bf}$ and $w_{\rm bf'}$ on the $\rm bf$ and $\rm bf'$ sides, respectively) and the displacement of the line with respect to the body ($\delta X_{\rm bf} = \phi_{0,{\rm bf}} \cdot \delta X_0$ and $\delta X_{\rm bf'} = \phi_{0,{\rm bf'}} \cdot \delta X_0$ on the $\rm bf$ and $\rm bf'$ sides, respectively; here, $\delta X$ and $\delta X_0$ are considered as vectors, not necessarily normal to $\rm L$ and $\rm L_0$, respectively; $\phi_0$ defined in \ref{secEulLag}), i.e., $\delta X = w_{\rm bf} + \delta X_{\rm bf} = w_{\rm bf'} + \delta X_{\rm bf'}$ (see Fig.~\ref{deltaX}), this condition becomes
\begin{eqnarray}
&&- \int_{\rm L} (\sigma_{\rm s,bf} \cdot \nu_{\rm bf} + \sigma_{\rm s,bf'} \cdot \nu_{\rm bf'}
+ \gamma_{\rm ff'}\,\nu_{\rm ff'}) \cdot \delta X\,dl \nonumber\\
&&+ \int_{\rm L} ((\sigma_{\rm s,bf} \cdot \nu_{\rm bf}) \cdot \delta X_{\rm bf}
+ (\sigma_{\rm s,bf'} \cdot \nu_{\rm bf'}) \cdot \delta X_{\rm bf'})\,dl \nonumber\\
&&- \int_{\rm L} (\gamma_{\rm bf}\,\nu_{\rm bf} \cdot \delta X_{\rm bf}
+ \gamma_{\rm bf'}\,\nu_{\rm bf'} \cdot \delta X_{\rm bf'})\,dl = 0 \label{bff'variational}
\end{eqnarray}
(the last term of the condition being written in Eulerian form, using $\gamma\,da = \gamma_0\,da_0$ for $\rm bf$ and $\rm bf'$), which leads to two equilibrium equations at the contact line (as in the case of the thin plate: \cite{Olives:1993,Olives:1996}).
The first one
\begin{eqnarray}
\sigma_{\rm s,bf} \cdot \nu_{\rm bf} + \sigma_{\rm s,bf'} \cdot \nu_{\rm bf'}
+ \gamma_{\rm ff'}\,\nu_{\rm ff'} = 0 \label{bff'equilibrium1}
\end{eqnarray}
corresponds to a contact line fixed on the body ($\delta X_0 = 0$, hence $\delta X_{\rm bf} = \delta X_{\rm bf'} = 0$) and expresses the equilibrium of the two surface stresses and the fluid--fluid surface tension. This shows that the surface stresses are forces acting on a line fixed to the material points of the body. Note that this equation suggests that a finite displacement occurs at the contact line (in the next section, an explicit example of finite-displacement solution will be presented). Some authors \citep{Madasu-Cairncross:2004} proposed the presence at the contact line of a force originating from the volume stresses $\sigma$ in the body. The preceding equation shows that there is no such volume stress contribution. This is a consequence of the validity of Green's formula, as mentioned above, and will be illustrated in the next section (Subsec.~\ref{secvalidGreen}). With the help of (\ref{bff'equilibrium1}), the above line condition gives the second equation (according to $\delta X_{\rm bf} = \phi_{0,{\rm bf}} \cdot \delta X_0$ and $\delta X_{\rm bf'} = \phi_{0,{\rm bf'}} \cdot \delta X_0$)
\begin{eqnarray}
\phi_{0,{\rm bf}}^* \cdot (\sigma_{\rm s,bf} - \gamma_{\rm bf}\,I) \cdot \nu_{\rm bf} 
+ \phi_{0,{\rm bf'}}^* \cdot (\sigma_{\rm s,bf'} - \gamma_{\rm bf'}\,I') \cdot \nu_{\rm bf'} = 0 \label{bff'equilibrium20}
\end{eqnarray}
($I$ and $I'$ are the identity mappings on ${\rm T}_x(\rm S_{bf})$ and ${\rm T}_x(\rm S_{bf'})$, respectively), which corresponds to a line moving on the body ($\delta X_0 \neq 0$), i.e., with $\phi_{\rm r} = \phi_{0,{\rm bf'}} \cdot \phi_{0,{\rm bf}}^{-1}$ (which is the `relative deformation gradient' of the $\rm bf'$ side with respect to the $\rm bf$ side; this concept was defined in \cite{Olives-Bronner:1984}; note that $\phi_{\rm r}$ does not depend on the reference state: \cite{Olives:2010a}),
\begin{eqnarray}
(\sigma_{\rm s,bf} - \gamma_{\rm bf}\,I) \cdot \nu_{\rm bf} 
+ \phi_{\rm r}^* \cdot (\sigma_{\rm s,bf'} - \gamma_{\rm bf'}\,I') \cdot \nu_{\rm bf'} 
= 0.\label{bff'equilibrium2}
\end{eqnarray}
This equation expresses the equilibrium of the forces acting on the `free' contact line (not fixed to the material points of the body). In the reference state, these forces (normal to the line and positively measured from $\rm bf$ to $\rm bf'$) are represented by the opposite of the first member of Eq. (\ref{bff'equilibrium20}). Applying $\tau^*$ to the last equation (where $\tau$ is a unit vector tangent to the contact line at $x$; thus, $\phi_{\rm r} \cdot \tau = \tau$) gives the same equation as the tangential component (along $\tau$) of (\ref{bff'equilibrium1}). Applying $\nu_{\rm bf}^*$ to (\ref{bff'equilibrium2}) leads to
\begin{eqnarray}
\sigma_{\rm bf,\nu\nu} - \gamma_{\rm bf}
- (\sigma_{\rm bf',\nu\nu} - \gamma_{\rm bf'})\,a_{\rm r,\nu\nu}
+ \sigma_{\rm bf',\tau\nu}\,a_{\rm r,\tau\nu} = 0, \label{bff'equilibrium2a}
\end{eqnarray}
where $(\sigma_{\rm bf,\nu\nu}, \sigma_{\rm bf,\tau\nu})$ are the components of $\sigma_{\rm s,bf} \cdot \nu_{\rm bf}$ in the basis $(\nu_{\rm bf}, \tau)$, similarly for $\sigma_{\rm s,bf'}$ with the basis $(\nu_{\rm bf'}, \tau)$, and $(a_{\rm r,\nu\nu}, a_{\rm r,\tau\nu})$ the components of $\phi_{\rm r} \cdot \nu_{\rm bf}$ in the basis $(-\nu_{\rm bf'}, \tau)$ (thus, $a_{\rm r,\nu\nu} > 0$). With the help of (\ref{bff'equilibrium1}), this equation may be written in the more geometrical form \citep{Olives:2010a}
\begin{eqnarray}
-\gamma_{\rm bf} + \gamma_{\rm bf'}\,a_{\rm r,\nu\nu}
+ \gamma_{\rm ff'}\,\frac{\sin\varphi_{\rm f'} - a_{\rm r,\nu\nu}\,\sin\varphi_{\rm f}}{\sin\varphi_{\rm b}}
+ \sigma_{\rm bf',\tau\nu}\,a_{\rm r,\tau\nu} = 0, \label{bff'equilibrium2b}
\end{eqnarray}
or
\begin{eqnarray}
&&-\gamma_{\rm bf} + \gamma_{\rm bf'}\,a_{\rm r,\nu\nu}- \gamma_{\rm ff'}\,\cos\varphi_{\rm f} \nonumber\\
&&- \gamma_{\rm ff'}\,\sin\varphi_{\rm f}\,
\frac{\cos\varphi_{\rm b} + a_{\rm r,\nu\nu}}{\sin\varphi_{\rm b}}
+ \sigma_{\rm bf',\tau\nu}\,a_{\rm r,\tau\nu} = 0,\label{bff'equilibrium2c} 
\end{eqnarray}
where $\varphi_{\rm f}$, $\varphi_{\rm f'}$ and $\varphi_{\rm b}$ are the three angles of contact, respectively measured in $\rm f$, $\rm f'$ and $\rm b$ ($\varphi_{\rm f} + \varphi_{\rm f'} + \varphi_{\rm b} = 2\pi$). This shows that the classical capillary Young's equation is strongly modified and replaced with the preceding one (as it occurred for the thin plate: \cite{Olives:1993,Olives:1996}). In the limit case of an undeformable solid, owing to $a_{\rm r,\nu\nu} = 1$, $a_{\rm r,\tau\nu} = 0$ and $\varphi_{\rm b} = \pi$ ($\lim_{\varphi_{\rm b} \rightarrow \pi} \frac{\cos\varphi_{\rm b} + 1}{\sin\varphi_{\rm b}} = 0$), this equation leads to the classical Young's equation $-\gamma_{\rm bf} + \gamma_{\rm bf'} - \gamma_{\rm ff'}\,\cos\varphi_{\rm f} = 0$. Note that, for an undeformable solid, (\ref{bff'equilibrium1}) cannot be obtained from the variational condition (\ref{bff'variational}) because, if the line is fixed on the body ($\delta X_0 = 0$), then $\delta X = 0$ (since $w_{\rm bf} = w_{\rm bf'} = 0$). In this case (in which $\delta X = \delta X_{\rm bf} = \delta X_{\rm bf'}$), (\ref{bff'variational}) leads to only one line equation, which is the classical Young's equation (equilibrium of forces acting on the `free' contact line). Note also that, if the deformable body $\rm b$ is a fluid, then $\sigma_{\rm s} = \gamma\,I$ for the $\rm bf$ and $\rm bf'$ fluid--fluid surfaces (as shown above, after (\ref{bfequilibrium1normal})), so that the second line equation (\ref{bff'equilibrium20}) is obviously satisfied, while the first one (\ref{bff'equilibrium1}) leads to the classical equilibrium of the three fluid--fluid surface tensions. Then, in this particular case too, there is only one line equation.

\section{Example of finite-displacement solution} \label{secsolution}

We have shown in the preceding section that the coherence of the theory is mainly based on the validity of Green's formula (\ref{Greenvolume}) despite the contact line singularity. This point is justified in the present section, with an explicit example of solution concerning the simple case of a half-space elastic solid, bounded by a plane, and subjected to a constant normal surface tension concentrated on a straight line of its surface. This solution satisfies the above line equations (\ref{bff'equilibrium1})--(\ref{bff'equilibrium2c}), its singularity at the contact line is described, its displacement field remains finite, the elastic energy is also finite, and it is shown that Green's formula remains valid at the contact line.

\subsection{A plane strain problem}

\begin{figure}[t]
\begin{center}
\includegraphics[height=4.5cm]{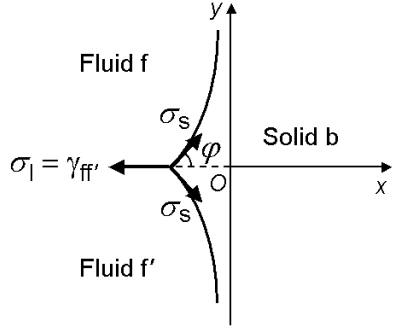}
\end{center}
\caption{Half-space elastic solid subjected to a normal force concentrated on a straight line of its surface. We present a solution with a finite displacement and the formation of an edge at this contact line.}\label{Solution}
\end{figure}

Let the body $\rm b$ be an isotropic elastic solid occupying (in the reference state) the half space $x \geq 0$, in the orthonormal frame $(Ox, Oy, Oz')$, with a constant and isotropic surface stress of eigenvalue $\sigma_{\rm s}$ on its surface $x = 0$, and subjected to a constant force (per unit length) $\sigma_{\rm l}$ parallel to $Ox$, concentrated on the line $x = y = 0$ (Fig.~\ref{Solution}; sign convention: $\sigma_{\rm l} > 0$ if the force is directed to the outside of $\rm b$; there is no gravity: $\bar g = 0$). Clearly, it is equivalent to consider that the body is in contact with a fluid $\rm f$ occupying the region $x < 0$ and $y > 0$ (in the reference state), and a fluid $\rm f'$ the other region $x < 0$ and $y < 0$, with $\gamma_{\rm ff'} = \sigma_{\rm l}$ and isotropic surface stresses with the same eigenvalue $\sigma_{\rm s,bf} = \sigma_{\rm s,bf'} = \sigma_{\rm s}$. In the present equilibrium state (after deformation), owing to the symmetry of the problem with respect to the plane $y = 0$, and if the surface energies $\gamma_{\rm bf}$ and $\gamma_{\rm bf'}$ are the same function of the surface strain tensor $e_{\rm s}$ (temperature and chemical potentials being constant), then the preceding equation (\ref{bff'equilibrium2b}) is satisfied ($a_{\rm r,\nu\nu} = 1$, $a_{\rm r,\tau\nu} = 0$, $\gamma_{\rm bf} = \gamma_{\rm bf'}$ and $\varphi_{\rm f} = \varphi_{\rm f'}$, by symmetry). The other equation (\ref{bff'equilibrium1}) at the contact line gives here
\begin{eqnarray}
\sigma_{\rm l} = 2 \sigma_{\rm s} \cos\varphi, \label{equline}
\end{eqnarray}
where $\varphi = \varphi_{\rm b}/2 = \pi - \varphi_{\rm f} = \pi - \varphi_{\rm f'}$, which determines the angle $\varphi$, i.e., the orientation of the vector $\nu_{\rm bf}$ tangent to the $\rm bf$ side of the surface (see Fig.~\ref{Solution}).
At the surface of the body, instead of applying the complex stress condition (\ref{bfequilibrium1}), we shall impose a simple displacement condition:
\begin{eqnarray}
u_x = \frac{-a}{|y| + b},\; u_y = u_{z'} = 0 \label{ux}
\end{eqnarray}
($a \neq 0$, $b > 0$), at any point $(0,y,z')$ of the surface. The value of $a/b^2$ is fixed by (\ref{equline}):
\begin{eqnarray}
\frac{a}{b^2} &=& \partial_y u_x(y = 0^+) = \frac{1}{\tan\varphi} \nonumber\\
&=& \frac{\rho}{\sqrt{1 - \rho^2}},\;\rm{with}\;\rho = \frac{\sigma_{\rm l}}{2 \sigma_{\rm s}}. \label{equlinebis}
\end{eqnarray}
In the following (Secs.~\ref{secFandG} and \ref{secsolsing}), we solve the problem with $b = 1$, in the frame of classical plane strain elasticity, i.e., with 
\begin{eqnarray}
&u_x&, u_y\;{\rm functions\;of}\;(x, y)\nonumber\\
&u_{z'}& = 0.\label{uxuyuz}
\end{eqnarray}
By a change of variables (Sec.~\ref{secchangevar}), this will lead to solutions for any $a$ and $b$ satisfying (\ref{equlinebis}).

\subsection{The analytic functions $F$ and $G$} \label{secFandG}

In the following, $z$ will denote the complex variable $x + iy$ and $u$ the complex displacement $u_x + iu_y$ (function of the complex variable $z$). We use the general Kolosov's solution of plane strain elasticity
\begin{eqnarray}
u(z) &=& - \frac{1}{2\mu}\,(k\,F(z) + z\,\overline{F'(z)} + \overline{G(z)})\nonumber\\
{\rm where}\quad k &=& - \frac{\lambda + 3\mu}{\lambda + \mu}\label{displacementFG}
\end{eqnarray}
($\lambda$, $\mu > 0$ Lam\'e's coefficients, $-3 < k < -1$), based on the two analytic functions $F$ and $G$. We then follow Muskhelishvili's method (e.g.~\cite{Mandel:1966}, tome II, annexe XVI)---adapted to the present singularity problem---to determine $F$ and $G$. The mapping $\displaystyle \zeta \rightarrow z = \omega (\zeta) = \frac{1 - \zeta}{1 + \zeta}$ from ${\bf C} - \{-1\}$ onto itself is bijective, analytic and $\omega^{-1} = \omega$. It transforms ${\rm B} = \{\zeta \in {\bf C}|\;|\zeta| < 1\}$ into ${\rm A} = \{z \in {\bf C}|\;\Re\,z > 0\}$, and ${\bf U} - \{-1\}$ into ${\rm D} = \{z \in {\bf C}|\;\Re\,z = 0\}$ (where ${\bf U} = \{\zeta \in {\bf C}|\;|\zeta| = 1\}$). The above displacement condition, with $b = 1$, on the surface of the body thus means
\begin{eqnarray}
&&k\,F(z_0) + z_0\,\overline{F'(z_0)} + \overline{G(z_0)} = f(z_0)
\quad{\rm for}\;z_0 \in {\rm D},\nonumber\\
&&{\rm where}\;f(z_0) = \frac{1}{|y_0| + 1},\quad y_0 = \Im\,z_0, \label{conditionz}
\end{eqnarray}
i.e., with the variable $\zeta$
\begin{eqnarray}
k\,\Phi(\zeta_0) + \frac{\omega(\zeta_0)}{\overline{\omega'(\zeta_0)}}\,\overline{\Phi'(\zeta_0)} + \overline{\Psi(\zeta_0)} = \phi(\zeta_0)\quad{\rm for}\;\zeta_0 \in {\bf U} \label{conditionzeta1}
\end{eqnarray}
(extended to $\zeta_0 = -1$), where $\Phi(\zeta) = F(\omega (\zeta))$, $\Psi(\zeta) = G(\omega (\zeta))$ and
\begin{eqnarray}
\phi(\zeta_0) &=& f(z_0) = \frac{1}{|y_0| + 1} = \frac{1}{i\varepsilon \frac{1 - \zeta_0}{1 + \zeta_0} + 1}\nonumber\\
&=& \frac{1 + \zeta_0}{1 + \zeta_0 + i\varepsilon (1 - \zeta_0)},\label{phi}
\end{eqnarray}
where $\varepsilon = {\rm sign}(\Im\,\zeta_0)$, for any $\zeta_0 \in {\bf U}$ (since $|y_0| = -i z_0\,{\rm sign}\,y_0 = i\varepsilon z_0$). The function $\displaystyle \theta(\zeta) = \frac{\overline{\omega(\zeta)}}{\omega'(\zeta)} = \frac{1 - \bar\zeta}{1 + \bar\zeta}\cdot\frac{(1 + \zeta)^2}{-2}$ is not analytic, but its restriction to ${\bf U}$
\begin{eqnarray*}
\theta(\zeta_0) = \frac{1 - \frac{1}{\zeta_0}}{1 + \frac{1}{\zeta_0}}\cdot\frac{(1 + \zeta_0)^2}{-2}
= \frac{1 - \zeta_0^2}{2}
\end{eqnarray*}
is that of the analytic function $\displaystyle \chi(\zeta) = \frac{1 - \zeta^2}{2}$. Since (\ref{conditionzeta1}) may be written as
\begin{eqnarray}
k\,\Phi(\zeta_0) + \overline{\Xi(\zeta_0)} = \phi(\zeta_0)\quad{\rm for}\;\zeta_0 \in {\bf U}, \label{conditionzeta2}
\end{eqnarray}
in which $\Xi = \chi\,\Phi' + \Psi$ is analytic, we propose to {\it define} $\Phi$ by
\begin{eqnarray}
k\,\Phi(\zeta) = \frac{1}{2\pi i} \int_\gamma \frac{\phi(\zeta_0)}{\zeta_0 - \zeta}\,d\zeta_0 + C
\quad{\rm for}\;\zeta \in {\rm B}
\end{eqnarray}
(see \cite{Mandel:1966}, ibid.), where $C$ is a constant and $\gamma$ the circuit $t \in [0,2\pi] \rightarrow e^{it}$. Since $\phi$ is continuous in $\bf U$, $\Phi$ is analytic in $\rm B$. Using the decomposition
\begin{eqnarray*}
\frac{\phi(\zeta_0)}{\zeta_0 - \zeta} &=& 
\frac{1 + \zeta_0}{(1 + \zeta_0 + i\varepsilon (1 - \zeta_0))(\zeta_0 - \zeta)}
= \frac{1 + \zeta_0}{(1 - i\varepsilon)(\zeta_0 + i\varepsilon)(\zeta_0 - \zeta)}\\
&=& \frac{-1}{i\varepsilon+\zeta}\cdot\frac{1}{\zeta_0+i\varepsilon} 
+ \frac{(1+i\varepsilon)(1+\zeta)}{2(i\varepsilon+\zeta)}\cdot\frac{1}{\zeta_0-\zeta},
\end{eqnarray*}
and according to
\begin{eqnarray*}
\int_{\gamma^+} \frac{d\zeta_0}{\zeta_0 + i} = \int_{\gamma^-} \frac{d\zeta_0}{\zeta_0 - i} = \frac{\pi}{2}i\\
\int_{\gamma^+} \frac{d\zeta_0}{\zeta_0 - \zeta} + \int_{\gamma^-} \frac{d\zeta_0}{\zeta_0 - \zeta} = 2\pi i 
\end{eqnarray*}
($\zeta \in \rm B$; $\gamma^+:t \in [0,\pi] \rightarrow e^{it}$ and $\gamma^-:t \in [\pi,2\pi] \rightarrow e^{it}$) and
\begin{eqnarray*}
&&\int_{\gamma^+} \frac{d\zeta_0}{\zeta_0 - \zeta} - \int_{\gamma^-} \frac{d\zeta_0}{\zeta_0 - \zeta}\\
&=& -\int_{\beta^+} \frac{d\zeta_0}{\zeta_0 - \zeta} - \int_{\alpha} \frac{d\zeta_0}{\zeta_0 - \zeta}
- \int_{\beta^-} \frac{d\zeta_0}{\zeta_0 - \zeta} - \int_{\alpha} \frac{d\zeta_0}{\zeta_0 - \zeta}\\
&=& \pi i - \int_{\alpha} \frac{d\zeta_0}{\zeta_0 - \zeta}
-\pi i - \int_{\alpha} \frac{d\zeta_0}{\zeta_0 - \zeta} = -2 \int_{\alpha} \frac{d\zeta_0}{\zeta_0 - \zeta}\\
&=& -2 \int_{\alpha_1} \frac{dz}{z} = -2 \log \frac{1 - \zeta}{1 + \zeta},
\end{eqnarray*}
where $\beta^+$ is the path $t \in [0,\pi] \rightarrow e^{i(\pi-t)}$ followed by the path $t \in [0,1] \rightarrow 1+2t\zeta$, $\beta^-$ the path $t \in [\pi,2\pi] \rightarrow e^{it}$ followed by the path $t \in [0,1] \rightarrow 1+2t\zeta$, $\alpha$ the path $t \in [0,1] \rightarrow 1+2(1-t)\zeta$, $\alpha_1$ the path $t \in [0,1] \rightarrow 1+\zeta-2t\zeta$, and $\log$ defined in $\bf C - R_-$, we finally obtain
\begin{eqnarray}
k\,\Phi(\zeta) = \frac{1}{2(1 + \zeta^2)} (1 + \zeta + \zeta^2 
+ \frac{1 - \zeta^2}{\pi} \log \frac{1 - \zeta}{1 + \zeta}) + C
\quad{\rm for}\;\zeta \in {\rm B},\label{Phi}
\end{eqnarray}
i.e., with the variable $z = \omega(\zeta)$
\begin{eqnarray}
k\,F(z) &=& \frac{1}{4(1 + z^2)} (3 + z^2 + \frac{4}{\pi} z\log z) + C\nonumber\\
&=& \frac{\frac{\pi}{2} + z\log z}{\pi(1 + z^2)},\label{F}
\end{eqnarray}
with $C = -\frac{1}{4}$. Since $\frac{\pi}{2} + z\log z = (z - i)F_1(z)$ ($(z + i)F_1(z)$, respectively) and $\frac{1}{1 + z^2} = \frac{1}{z - i} F_2(z)$ ($\frac{1}{z + i} F_2(z)$, respectively), where $F_1$ and $F_2$ are analytic in a neighbourhood of $i$ (of $-i$, respectively), the function $F$ is analytic in $\bf C - R_-$, and then in $\rm A$. According to $\lim_{z \rightarrow 0} k\,F(z) = \frac{1}{2}$, it may be extended as a continuous function in $\bf C - R^*_-$, and then in $\bar{\rm A} = \rm A \cup D$. We may write
\begin{eqnarray}
k\pi\,F(z) &=& z\log z + g(z),\;{\rm i.e.}\nonumber\\
(1 + z^2)g(z) &=& \frac{\pi}{2} - z^3\log z,\label{g}
\end{eqnarray}
where $g$ may be continuously extended at $0$, and after derivation
\begin{eqnarray}
k\pi\,F'(z) &=& \log z + 1 + g'(z),\nonumber\\
2z\,g(z) + (1 + z^2)g'(z) &=& - 3z^2\log z - z^2,\label{g'}
\end{eqnarray}
this last equality showing that $g'$ may be continuously extended at $0$. We know that $F'$ is analytic in $\bf C - R_-$ (and then continuous in $\bar{\rm A} - \{0\}$) and (\ref{g'}) shows that $z\,F'(z)$ and ${\bar z}\,F'(z)$ may be extended as continuous functions in $\bf C - R^*_-$ (and then in $\bar{\rm A}$).

Since $\overline{F(z_0)} = F(\bar{z_0}) = F(-z_0)$ for $z_0 \in \rm D$,
\begin{eqnarray}
k\,F(z_0) + k\,\overline{F(z_0)} 
&=& \frac{\pi + z_0\log z_0 - z_0\log (-z_0)}{\pi(1 + z_0^2)}\nonumber\\
&=& \frac{\pi + z_0(-i \varepsilon \pi)}{\pi(1 + z_0^2)} = \frac{1 - |y_0|}{1 - y_0^2}\nonumber\\
&=& \frac{1}{1 + |y_0|} = f(z_0)\label{kFkFbar}
\end{eqnarray}
(with the notations of (\ref{conditionz}) and (\ref{phi})), so that (\ref{conditionz}) gives
\begin{eqnarray}
G(z_0) &=& f(z_0) - k\,\overline{F(z_0)} - \bar{z_0}\,F'(z_0)\nonumber\\
&=& k\,F(z_0) + z_0\,F'(z_0)\quad{\rm for}\;z_0 \in {\rm D}.
\end{eqnarray}
Owing to this expression, we then {\it define} $G$ in $\bf C - R^*_-$ as
\begin{eqnarray}
G(z) = k\,F(z) + z\,F'(z),\label{G}
\end{eqnarray}
which is analytic in $\bf C - R_-$ (and then in $\rm A$) and continuous in $\bf C - R^*_-$ (and then in $\bar{\rm A}$).

\subsection{The solution and its singularity} \label{secsolsing}

According to (\ref{displacementFG}) and (\ref{G}), our solution $u$ is then 
\begin{eqnarray}
u(z) = - \frac{1}{2\mu}(k(F(z) + \overline{F(z)}) + (z + \bar z)\overline{F'(z)}),
\label{displacementF}
\end{eqnarray}
with $F$ given by (\ref{F}), and is a continuous function in $\bar{\rm A}$. Its value, for $z_0 \in \rm D$,
\begin{eqnarray}
u(z_0) = - \frac{1}{2\mu}k(F(z_0) + \overline{F(z_0)}) 
= - \frac{1}{2\mu}\cdot\frac{1}{1 + |y_0|}\label{uD}
\end{eqnarray}
(from (\ref{kFkFbar})) has the form (\ref{ux}) with $b = 1$. The displacement $u$ is then finite at $z = 0$ (i.e., at the contact line).

When the variable $\zeta = \omega^{-1}(z)$ tends to $-1$, $|z|$ tends to $+\infty$. Using this variable and the expression (\ref{Phi}) (with $C = -\frac{1}{4}$), we have
\begin{eqnarray*}
-2\mu\,\overline{u(z)} &=& k(F(z) + \overline{F(z)}) + (z + \bar z)F'(z)\\
&=& 2\Re(k\,\Phi(\zeta)) 
+ (\frac{1 - \zeta}{1 + \zeta} + \frac{1 - \bar\zeta}{1 + \bar\zeta})\frac{(1+\zeta)^2 \Phi'(\zeta)}{-2}\\
&=& 2\Re(k\,\Phi(\zeta))\\
&-&\frac{(1+\zeta)^2}{|1+\zeta|^2}\cdot\frac{1-|\zeta|^2}{k\pi(1+\zeta^2)^2}
(\frac{\pi}{2}(1-\zeta^2) - (1+\zeta^2) - 2\zeta\log\frac{1 - \zeta}{1 + \zeta}),
\end{eqnarray*}
which tends to $0$ when $\zeta$ tends to $-1$ ($\Phi(\zeta)$ tends to $0$ owing to $(1-\zeta^2)\log\frac{1 - \zeta}{1 + \zeta} = (1-\zeta^2)\log(1 - \zeta) - (1 - \zeta)(1 + \zeta)\log(1 + \zeta)$; similarly, $(1-|\zeta|^2)\log\frac{1 - \zeta}{1 + \zeta} = (1-|\zeta|^2)\log(1 - \zeta) - (1 - |\zeta|^2)\log(1 + \zeta) = (1-|\zeta|^2)\log(1 - \zeta) - \frac{1}{2}(1 - \bar\zeta)(1 + \zeta)\log(1 + \zeta) - \frac{1}{2}(1 - \zeta)(1 + \bar\zeta)\log(1 + \zeta)$ also tends to $0$). This shows that the displacement $u(z)$ tends to $0$ when $|z|$ tends to $+\infty$.

Kolosov's expressions of the strain and stress tensors components are then obtained from (\ref{displacementFG}) and, according to (\ref{G}),
\begin{eqnarray}
\varepsilon_{xx} &=& \frac{1}{2\mu}\Re(-(1 + k)F'(z) - \bar z\,F''(z) - G'(z))\nonumber\\
&=& \frac{1}{2\mu}\Re(-2(1 + k)F'(z) - (z + \bar z)F''(z)),\nonumber\\
\varepsilon_{yy} &=& \frac{1}{2\mu}\Re(-(1 + k)F'(z) + \bar z\,F''(z) + G'(z))\nonumber\\
&=& \frac{1}{2\mu}\Re((z + \bar z)F''(z)),\nonumber\\
\varepsilon_{xy} &=& \frac{1}{2\mu}\Im(\bar z\,F''(z) + G'(z))\nonumber\\
&=& \frac{1}{2\mu}\Im((1 + k)F'(z) + (z + \bar z)F''(z)),\label{epsilonij}\\
\sigma_{xx} &=& \Re(2\,F'(z) - \bar z\,F''(z) - G'(z))\nonumber\\
&=& \Re((1 - k)F'(z) - (z + \bar z)F''(z)),\nonumber\\
\sigma_{yy} &=& \Re(2\,F'(z)  + \bar z\,F''(z) + G'(z))\nonumber\\
&=& \Re((3 + k)F'(z) + (z + \bar z)F''(z)),\nonumber\\
\sigma_{xy} &=& \Im(\bar z\,F''(z) + G'(z))\nonumber\\
&=& \Im((1 + k)F'(z) + (z + \bar z)F''(z)),\nonumber\\
\sigma_{z'z'} &=& \frac{2\lambda}{\lambda + \mu}\Re(F'(z)) = (3 + k)\Re(F'(z)).\label{sigmaij}
\end{eqnarray}

By derivation of (\ref{g'}),
\begin{eqnarray}
k\pi\,F''(z) &=& \frac{1}{z} + g''(z),\nonumber\\
2\,g(z) + 4z\,g'(z) + (1 + z^2)g''(z) &=& - 6z\log z - 5z,\label{g''}
\end{eqnarray}
the last equality showing that $g''$ may be continuously extended at $0$. The function $F''$ is analytic in $\bf C - R_-$ (and then continuous in $\bar{\rm A} - \{0\}$) and (\ref{g''}) shows that $z\,F''(z)$ may be extended as a continuous function in $\bf C - R^*_-$ (and then in $\bar{\rm A}$) and that
\begin{eqnarray}
\lim_{z \rightarrow 0,\;\theta\;\rm{constant}} (z +{\bar z})F''(z) = \frac{1 + e^{-2i\theta}}{k\pi},\label{limF''}
\end{eqnarray}
where $\theta = \arg z$. Let $r_0$ be $> 0$ and ${\rm V_0} = \{z \in {\bf C}|\;\Re\,z > 0\;{\rm and}\;|z| < r_0\}$. The function $z\,F''(z)$ is continuous and then bounded in the compact set $\overline{\rm V_0}$, so that $(z +{\bar z})F''(z)$ is also bounded in $\overline{\rm V_0} - \{0\}$. In addition, the first equality (\ref{g'}), where $g'$ is continuous in $\bar{\rm A}$ and then bounded in $\overline{\rm V_0}$, and $g'(0) = 0$ (consequence of the second equality (\ref{g'})) lead to
\begin{eqnarray}
&\lim_{z \rightarrow 0}& \Re(F'(z)) = + \infty \nonumber\\
&\lim_{z \rightarrow 0,\;\theta\;\rm{constant}}& \Im(F'(z)) = \frac{\theta}{k\pi}\nonumber\\
&|\Re(F'(z))|& \leq c_1 |\log r| + d_1\quad{\rm in}\;\overline{\rm V_0} - \{0\}\nonumber\\
&\Im(F'(z))&\;{\rm is\;bounded\;in}\;\overline{\rm V_0} - \{0\}\label{limF'}
\end{eqnarray}
($r = |z|$, $\theta = \arg z$; $c_1$, $d_1$ constants $> 0$).

The expressions (\ref{epsilonij}) and (\ref{sigmaij}) and the preceding results show that all the components of the strain and stress tensors are continuous in $\bar{\rm A} - \{0\}$, bounded in $\overline{\rm V_0} - \{0\}$ excepted
\begin{eqnarray}
|\varepsilon_{xx}|, |\sigma_{xx}|, |\sigma_{yy}|\;{\rm and}\;|\sigma_{z'z'}| \leq c |\log r| + d
\quad{\rm in}\;\overline{\rm V_0} - \{0\}\label{boundstrainstress}
\end{eqnarray}
(different constants $c$, $d$ $> 0$ for each strain or stress tensor component), and
\begin{eqnarray}
&\lim_{z \rightarrow 0}& \varepsilon_{xx} = + \infty \nonumber\\
&\lim_{z \rightarrow 0,\;\theta\;\rm{constant}}& \varepsilon_{yy} 
= \frac{1}{2\mu}\cdot\frac{1 + \cos 2\theta}{k\pi}\nonumber\\
&\lim_{z \rightarrow 0,\;\theta\;\rm{constant}}& \varepsilon_{xy} 
= \frac{1}{2\mu}\cdot\frac{(1 + k)\theta - \sin 2\theta}{k\pi}\nonumber\\
&\lim_{z \rightarrow 0}& \sigma_{xx},\;\sigma_{yy}\;{\rm and}\;\sigma_{z'z'} = + \infty\nonumber\\
&\lim_{z \rightarrow 0,\;\theta\;\rm{constant}}& \sigma_{xy} 
= \frac{(1 + k)\theta - \sin 2\theta}{k\pi}.
\end{eqnarray}

Explicit expressions of $F'$ and $F''$ to be used in (\ref{epsilonij}) and (\ref{sigmaij}) are obtained from (\ref{F}): 
\begin{eqnarray}
k\pi\,F'(z) &=& \frac{1 - \pi z + z^2 + (1 - z^2)\log z}{(1 + z^2)^2}\nonumber\\
k\pi\,F''(z) &=& \frac{1 - \pi z - 2z^2 + 3\pi z^3 - 3z^4 + (-6z^2 + 2z^4)\log z}{z(1 + z^2)^3}.\label{F'F''}
\end{eqnarray}

\subsection{Change of variables} \label{secchangevar}

As noted after (\ref{equlinebis}), we have solved the problem with the value $b = 1$, which means (physical) dimensionless quantities $y = \Re z$ and $z$, and, according to (\ref{conditionz}), dimensionless quantities $F(z)$ and $G(z)$. Physical quantities are obtained by considering the new variable $z/b$ and the new functions $\widetilde{F}(z) = a' F(z/b)$ and $\widetilde{G}(z) = a' G(z/b) = k\,\widetilde{F}(z) + z\,\widetilde{F}'(z)$, where $b > 0$ is a length and $a' \in {\bf R^*}$ a force per unit length. With these new functions, the displacement $u$ given by (\ref{displacementFG}) or (\ref{displacementF}) becomes
\begin{eqnarray}
\tilde{u}(z) = a'\,u(\frac{z}{b})
\end{eqnarray}
and the components of the strain and stress tensors, $\varepsilon_{ij}$ and $\sigma_{ij}$ given by (\ref{epsilonij}) and (\ref{sigmaij}), become
\begin{eqnarray}
\tilde{\varepsilon}_{ij}(z) &=& \frac{a'}{b}\,\varepsilon_{ij}(\frac{z}{b})\nonumber\\
\tilde{\sigma}_{ij}(z) &=& \frac{a'}{b}\,\sigma_{ij}(\frac{z}{b}).
\end{eqnarray}
For $z_0 \in {\rm D}$, the displacement becomes
\begin{eqnarray}
\tilde{u}(z_0) = a'\,u(\frac{z_0}{b}) = - \frac{a'b}{2\mu}\cdot\frac{1}{|y_0| + b}
\end{eqnarray}
(from (\ref{uD})), which has the general form (\ref{ux}) with $a = \frac{a'b}{2\mu}$.

\subsection{Finite elastic energy} \label{secfiniteenergy}

The elastic energy per unit volume is
\begin{eqnarray}
\frac{\lambda}{2}(\varepsilon_{xx}+\varepsilon_{yy})^2 
+ \mu(\varepsilon_{xx}^2+\varepsilon_{yy}^2+2\varepsilon_{xy}^2)\label{energyvolume}
\end{eqnarray}
(using $\varepsilon_{ij}$ or $\tilde{\varepsilon}_{ij}$). Since $\varepsilon_{yy}$ and $\varepsilon_{xy}$ are continuous and bounded in $\overline{\rm V_0} - \{0\}$, $\varepsilon_{yy}^2$ and $\varepsilon_{xy}^2$ are integrable in ${\rm V_0}$ (considered as $\subset {\bf R}^2$). Owing to the inequality (\ref{boundstrainstress}), $\varepsilon_{xx}$ and $\varepsilon_{xx}^2$ are also integrable in ${\rm V_0}$, which finally implies that (\ref{energyvolume}) is integrable in ${\rm V_0}$. The elastic energy in ${\rm V_0}$
\begin{eqnarray}
E_{\rm el} = \int_{\rm V_0} (\frac{\lambda}{2}(\varepsilon_{xx}+\varepsilon_{yy})^2 
+ \mu(\varepsilon_{xx}^2+\varepsilon_{yy}^2+2\varepsilon_{xy}^2)) dx\,dy
\end{eqnarray}
(per unit length along the normal $Oz'$ to the $xy$ plane) is then well defined and finite.

\subsection{Validity of Green's formula} \label{secvalidGreen}

The assumption made in Sec.~\ref{secsurfaceline} that Green's formula (\ref{Greenvolume}) remains valid, in order to obtain the equilibrium equations at the triple contact line, will be now justified using our present finite-displacement solution. Since $u$ is continuous in $\bar{\rm A} \times {\bf R}$ (considered as $\subset {\bf R}^3$; we return here to the three-dimensional space, according to (\ref{uxuyuz})), we consider its variation $w = \delta u$ as also continuous in $\bar{\rm A} \times {\bf R}$ and then bounded in $\bar{\rm V}$, where ${\rm V} = {\rm V_0} \times ]0, l_0[$ ($l_0 > 0$), so that the components $w_i \in L^\infty({\rm V})$ ($\subset L^2({\rm V})$, since $\rm V$ is bounded). The partial derivatives $\partial_j u_i$ are $\partial_x u_x = \varepsilon_{xx}$, $\partial_y u_y = \varepsilon_{yy}$ (written in (\ref{epsilonij})) and
\begin{eqnarray}
\partial_y u_x &=& \frac{1}{2\mu}\Im(-(1 - k)F'(z) + \bar z\,F''(z) + G'(z))\nonumber\\
&=& \frac{1}{2\mu}\Im(2k\,F'(z) + (z + \bar z)F''(z)),\nonumber\\
\partial_x u_y &=& \frac{1}{2\mu}\Im((1 - k)F'(z) + \bar z\,F''(z) + G'(z))\nonumber\\
&=& \frac{1}{2\mu}\Im(2\,F'(z) + (z + \bar z)F''(z))\label{djui}
\end{eqnarray}
(obtained from (\ref{displacementFG}) and (\ref{G}); $\partial_j u_i = 0$ if either $i$ or $j$ refers to the third coordinate $z'$) and are all continuous in $(\bar{\rm A} - \{0\}) \times {\bf R}$, and bounded in $(\overline{\rm V_0} - \{0\}) \times {\bf R}$ excepted $\partial_x u_x = \varepsilon_{xx}$ which satisfies the inequality (\ref{boundstrainstress}) in $(\overline{\rm V_0} - \{0\}) \times {\bf R}$. We may then consider their variation $\delta(\partial_j u_i) = \partial_j w_i$ as also continuous in $(\bar{\rm A} - \{0\}) \times {\bf R}$, and bounded in $(\overline{\rm V_0} - \{0\}) \times {\bf R}$ excepted $\partial_x w_x$ which will satisfy an inequality similar to (\ref{boundstrainstress}) in $(\overline{\rm V_0} - \{0\}) \times {\bf R}$, so that all the derivatives $\partial_j w_i \in L^2({\rm V})$, then the components $w_i \in H^1({\rm V})$.

Similarly, the components of the stress tensor $\sigma_{ij} \in L^2({\rm V})$, since they are continuous in $(\bar{\rm A} - \{0\}) \times {\bf R}$, and either bounded in $(\overline{\rm V_0} - \{0\}) \times {\bf R}$ or satisfying the inequality (\ref{boundstrainstress}) in $(\overline{\rm V_0} - \{0\}) \times {\bf R}$. Since they are the real or imaginary part of a linear combination of $F'(z)$ and $(z + \bar z)F''(z)$ (see (\ref{sigmaij})), their partial derivatives $\partial_l \sigma_{ij}$ ($\partial_l = \partial_x$ or $\partial_y$) will have the form
\begin{eqnarray}
\partial_l\sigma_{ij} = \Re\;{\rm or}\;\Im\,(k_1\,F''(z) + k_2 (z + \bar z)F'''(z))\label{dlsigmaij}
\end{eqnarray}
(different constants $k_1$, $k_2$ for each $l, i, j$).

The expression (\ref{g''}), where $g''$ is continuous in $\bar{\rm A}$ and then bounded in $\overline{\rm V_0}$, leads to
\begin{eqnarray}
|F''(z)|& \leq \frac{c_2}{r} + d_2\quad{\rm in}\;\overline{\rm V_0} - \{0\}\label{majF''}
\end{eqnarray}
($c_2$, $d_2$ constants $> 0$). The derivation of (\ref{g''}) gives
\begin{eqnarray}
k\pi\,F'''(z) &=& -\frac{1}{z^2} + g'''(z),\nonumber\\
6\,g'(z) + 6z\,g''(z) + (1 + z^2)g'''(z) &=& - 6\log z - 11,\label{g'''}
\end{eqnarray}
the last equality showing that $z\,g'''(z)$ and $\bar z\,g'''(z)$ may be continuously extended at $0$, and then considered as continuous in $\bar{\rm A}$, and then bounded in $\overline{\rm V_0}$. The expression (\ref{g'''}) then leads to
\begin{eqnarray}
|(z + \bar z)F'''(z)|& \leq \frac{c_3}{r} + d_3\quad{\rm in}\;\overline{\rm V_0} - \{0\}\label{majF'''}
\end{eqnarray}
($c_3$, $d_3$ constants $> 0$). The expression (\ref{dlsigmaij}) and the inequalities (\ref{majF''}) and (\ref{majF'''}) show that the partial derivatives $\partial_l \sigma_{ij} \in L^1({\rm V})$.

Nevertheless, these derivatives $\partial_l \sigma_{ij} \not\in L^2({\rm V})$ (for $l = x$ or $y$, and $ij$ = $xx$, $yy$, $xy$ or $z'z'$), so that $\sigma_{ij} \not\in H^1({\rm V})$. Let us take the example of $\partial_x \sigma_{xx} = -\Re((1 + k)F''(z) + (z + \bar z)F'''(z))$:
\begin{eqnarray*}
-k\pi\,\partial_x \sigma_{xx} = \Re(\frac{1 + k}{z} - \frac{z + \bar z}{z^2} + h(z))
\end{eqnarray*}
(using (\ref{g''}) and (\ref{g'''})), where $h(z) = (1 + k)g''(z) + (z + \bar z)g'''(z)$ is continuous and bounded in $\overline{\rm V_0}$. Thus, $\partial_x \sigma_{xx} \not\in L^2({\rm V})$ because
\begin{eqnarray*}
(\Re(\frac{1 + k}{z} - \frac{z + \bar z}{z^2}))^2 = \frac{(k\cos\theta - \cos3\theta)^2}{r^2}
\end{eqnarray*}
is not integrable in $\rm V_0$.

Since $\sigma_{ij} \not\in H^1({\rm V})$, Green's formula (\ref{Greenvolume}) cannot be directly applied on ${\rm V}$, as noted in Sec.~\ref{secsurfaceline}. Nevertheless, in the following, we will show that Green's formula remains valid in this case. The open set ${\rm V}$ is bounded by the surfaces ${\rm S} = \{z \in {\bf C}|\;z = iy,\;0 < y < r_0\} \times ]0, l_0[$, ${\rm S'} = \{z \in {\bf C}|\;z = iy,\;-r_0 < y < 0\} \times ]0, l_0[$ and $\Sigma = (\{z \in {\bf C}|\;\Re\,z > 0\;{\rm and}\;|z| = r_0\} \times ]0, l_0[) \cup ({\rm V}_0 \times \{0\}) \cup ({\rm V}_0 \times \{l_0\})$. Since the components of $\sigma$ and $w$ belong to $C^1(\overline{{\rm V}_\varepsilon})$, where $0 < \varepsilon < r_0$ and ${\rm V}_\varepsilon = \{z \in {\bf C}|\;\Re\,z > 0\;{\rm and}\;\varepsilon < |z| < r_0\} \times ]0, l_0[$, Green's formula may be applied on ${\rm V}_\varepsilon$ (with $w = 0$ on $\Sigma$)
\begin{eqnarray}
\int_{{\rm V}_\varepsilon} {\rm tr}(\sigma^* \cdot {\rm D}w)\,dv
= - \int_{{\rm V}_\varepsilon} {\rm div}(\sigma^*) \cdot w\,dv
- \int_{{\rm S}_\varepsilon \cup {\rm S'}_\varepsilon \cup {\rm C}_\varepsilon} (\sigma^* \cdot w) \cdot n\,da,
 \label{Greenvolumeepsilon}
\end{eqnarray}
in which ${\rm S}_\varepsilon = \{z \in {\bf C}|\;z = iy,\;\varepsilon < y < r_0\} \times ]0, l_0[$, ${\rm S'}_\varepsilon = \{z \in {\bf C}|\;z = iy,\;-r_0 < y < -\varepsilon\} \times ]0, l_0[$, ${\rm C}_\varepsilon = \{z \in {\bf C}|\;\Re\,z > 0\;{\rm and}\;|z| = \varepsilon\} \times ]0, l_0[$ and the unit normal vectors $n$ are directed to the inside of ${\rm V}_\varepsilon$. Since the components of $\sigma$ and ${\rm D}w$ belong to $L^2({\rm V})$, ${\rm tr}(\sigma^* \cdot {\rm D}w) \in L^1({\rm V})$ which implies that $\int_{{\rm V}_\varepsilon} {\rm tr}(\sigma^* \cdot {\rm D}w)\,dv$ tends to $\int_{\rm V} {\rm tr}(\sigma^* \cdot {\rm D}w)\,dv$ when $\varepsilon \rightarrow 0$. Since the components of ${\rm div}(\sigma^*)$ belong to $L^1({\rm V})$ (the partial derivatives $\partial_l \sigma_{ij} \in L^1({\rm V})$) and those of $w$ to $L^\infty({\rm V})$, ${\rm div}(\sigma^*) \cdot w \in L^1({\rm V})$ which again implies that $\int_{{\rm V}_\varepsilon} {\rm div}(\sigma^*) \cdot w\,dv$ tends to $\int_{\rm V} {\rm div}(\sigma^*) \cdot w\,dv$ when $\varepsilon \rightarrow 0$. Moreover, according to (\ref{boundstrainstress}) and the functions $\log r$ and $(\log r)^2$ being integrable in $[0, 1]$ (their respective primitives, $r\log r - r$ and $r(\log r)^2 - 2r\log r + 2r$, tend to $0$ when $r \rightarrow 0$), the components of $\sigma$ belong to $L^2({\rm S})$ and $L^2({\rm S'})$. Since the components of $w$ also belong to $L^2({\rm S}\cup{\rm S'})$ (they are continuous and bounded in $\bar{\rm V}$), $(\sigma^* \cdot w) \cdot n \in L^1({\rm S}\cup{\rm S'})$ which shows that $\int_{{\rm S}_\varepsilon \cup {\rm S'}_\varepsilon} (\sigma^* \cdot w) \cdot n\,da$ tends to $\int_{{\rm S}\cup {\rm S'}} (\sigma^* \cdot w) \cdot n\,da$ when $\varepsilon \rightarrow 0$. Finally, for $i$ and $j$ fixed, the inequality
\begin{eqnarray*}
|\int_{{\rm C}_\varepsilon} \sigma_{ij}\,w_i\,n_j\,da|
\leq (c|\log \varepsilon| + d)\,e\,\pi\,\varepsilon\,l_0
\end{eqnarray*}
(from (\ref{boundstrainstress}); $e$ constant, $|w_i| \leq e$ in $\bar{\rm V}$) shows that
\begin{eqnarray}
\lim_{\varepsilon \rightarrow 0} \int_{{\rm C}_\varepsilon} (\sigma^* \cdot w) \cdot n\,da = 0.
\label{linevolumestress}
\end{eqnarray}
The limit of (\ref{Greenvolumeepsilon}) when $\varepsilon \rightarrow 0$ is then
\begin{eqnarray}
\int_{\rm V} {\rm tr}(\sigma^* \cdot {\rm D}w)\,dv
= - \int_{\rm V} {\rm div}(\sigma^*) \cdot w\,dv
- \int_{{\rm S} \cup {\rm S'}} (\sigma^* \cdot w) \cdot n\,da,
 \label{Greenvolumelimit}
\end{eqnarray}
i.e. Green's formula on $\rm V$. Note that some authors \citep{Madasu-Cairncross:2004} proposed that the volume stresses produced a resultant force at the contact line. The result (\ref{linevolumestress}) expresses that there is no such contribution of the volume stresses at the contact line (see also the comment after (\ref{bff'equilibrium1})).

\section{Conclusions}

In this paper, which concerns the mechanical surface properties of a deformable body, the general surface and contact line equations are first deduced from a variational formulation (see \cite{Olives:2010a} for the physical aspects of the theory), by applying Green's formula both in the whole space and on the Riemannian surfaces. Despite the singularity at the triple contact line (due to the action of the fluid--fluid surface tension on the body), it is assumed that Green's formula remains valid in order to obtain the equations at this line. The explicit example of solution given in Sec.~\ref{secsolution} justifies this assumption. The equations (\ref{bfequilibrium1}) and (\ref{bfequilibrium2}) at the surfaces are similar to the Cauchy's equations for the volume, but involve a new definition of the divergence term as a true divergence with respect to the tensorial product of the covariant derivatives on the surface and the whole space (till now, this term was only defined by its scalar product with a constant vector). The normal (\ref{bfequilibrium1normal}) and tangent (\ref{bfequilibrium1tangent}) components of the divergence equation (\ref{bfequilibrium1}) are respectively a generalization of the classical Laplace's equation and the surface tension hydrostatic equilibrium for a fluid--fluid interface. Similar equations were written for elastic solids, e.g. in \cite{Gurtin-Murdoch:1975}, \cite{Leo-Sekerka:1989}, \cite{Gurtin-etal:1998} and \cite{Steinmann:2008}. Note that our thermodynamic approach is valid for any deformable body (such as a viscoelastic solid, a viscous fluid or any other one). There are {\it two} equations at the contact line, which represent: (i) the equilibrium of the forces acting on the line fixed to the material points of the body (\ref{bff'equilibrium1}) (equilibrium of the two surface stresses and the fluid--fluid surface tension); (ii) the equilibrium of the forces acting on the `free' contact line (\ref{bff'equilibrium20})--(\ref{bff'equilibrium2c}) (i.e., line moving with respect to the material points of the body), which leads to a strong modification of the classical capillary Young's equation (as in the case of the thin plate: \cite{Olives:1993,Olives:1996}). These two equations reduce to only one equation in the particular case of the undeformable solid (leading to the classical Young's equation) or that of three fluids in contact (leading to the classical equilibrium of the three surface tensions). Note that (\ref{bff'equilibrium1}) shows that surface stresses are forces which act on a line fixed to the material points of the body and that the fluid--fluid surface tension is equilibrated by the two surface stresses (and not by the volume stresses of the body). This equation (\ref{bff'equilibrium1}) suggests a finite displacement and the formation of an edge at the contact line, contrary to the infinite-displacement solution obtained from classical elasticity \citep{Shanahan-deGennes:1986,Shanahan:1986} in which surface properties (such as surface stresses) were not taken into account. As a simplified image, the body--fluid interface behaves as a tensile membrane, which undergoes a finite displacement when subjected to a force concentrated on a line. Experiments seem to confirm this idea \citep{Jerison-etal:2011} and an experimental support of the above Eq. (\ref{bff'equilibrium1}) (Eq. (42) of \cite{Olives:2010a}) was recently obtained \citep{Style-etal:2013}. The existence of such a finite-displacement solution is shown with the explicit example of Sec.~\ref{secsolution} satisfying the line equations (\ref{bff'equilibrium1})--(\ref{bff'equilibrium2c}). This elastic solution, based on the approaches of Kolosov and Muskhelishvili---adapted to the present singularity problem---and the theory of analytic functions, leads to a description of the singularity at the contact line. While the displacement components are continuous functions, their first partial derivatives and the strain tensor components are discontinuous, generally having different limits when approaching the contact line 
under 
different directions (Sec.~\ref{secsolsing} and (\ref{djui})). This solution also leads to a finite value of the elastic energy (Sec.~\ref{secfiniteenergy}), whereas this energy is infinite in the classical elastic solution \citep{Shanahan-deGennes:1986,Shanahan:1986}. Owing to the contact line singularity, the stress tensor components do not belong to the Sobolev space $H^1({\rm V})$. Although Green's formula cannot be directly applied in this case, it is shown in Sec.~\ref{secvalidGreen} that this formula remains valid. This result justifies the theory leading to the line equations (\ref{bff'equilibrium1})--(\ref{bff'equilibrium2c}). It also proves, according to (\ref{linevolumestress}), that there is no force contribution of the volume stresses at the contact line (contrary to what was proposed in \cite{Madasu-Cairncross:2004}). In fact, (\ref{linevolumestress}) shows that the validity of Green's formula is equivalent to the absence of contribution of the volume stresses at the contact line. In the presented finite-displacement solution, the validity of Green's formula is a consequence of the inequalities (\ref{boundstrainstress}) for $\sigma_{ij}$ and (\ref{majF''}) and (\ref{majF'''}) for $\partial_l \sigma_{ij}$. The importance of Green's formula and its validity for a wider class of functions will be presented in a future paper.

\appendix
\section{Eulerian and Lagrangian surface quantities} \label{secEulLag}

Let us denote by $x_0$ the position of a point of the body in the reference state, $x$ its position in the present state, $u = x - x_0$ its displacement between the reference state and the present state, $x'$ its position in the varied state, $w = x' - x = \delta x$ its displacement between the present state and the varied state, $\partial_{0i} = \displaystyle\frac{\partial}{\partial x_0^i}$ and  $\partial_i = \displaystyle\frac{\partial}{\partial x^i}$. Let $\rm S_{0,bf}$, $\rm S_{bf}$ and $\rm S'_{bf}$ respectively be the $\rm bf$ dividing surfaces in the reference state, the present state and the varied state, and $(x_0^\alpha)$ and $(x^\alpha)$ arbitrary curvilinear coordinates on $\rm S_{0,bf}$ and $\rm S_{bf}$, respectively, where Greek indices $\alpha$, $\beta$, $\gamma$,... belong to $\{1,2\}$ ($(x_0^\alpha)$ and $(x^\alpha)$ must be clearly distinguished from the three-dimensional Cartesian coordinates $(x_0^i)$ and $(x^i)$, respectively). The geometrical transformations such that $F_0 : x_0 \rightarrow x$, defined in the part of $\rm E$ occupied by the body $\rm b$, will now be restricted to the $\rm bf$ surfaces. We thus have the mappings $F_0 : x_0 \rightarrow x$ from $\rm S_{0,bf}$ to $\rm S_{bf}$, $F : x \rightarrow x'$ from $\rm S_{bf}$ to $\rm S'_{bf}$, $j_0 : x_0 \rightarrow x_0$ from $\rm S_{0,bf}$ to $\rm E$, $j : x \rightarrow x$ from $\rm S_{bf}$ to $\rm E$, $G_0 : x_0 \rightarrow u$ from $\rm S_{0,bf}$ to $\rm E$ and $G : x \rightarrow w$ from $\rm S_{bf}$ to $\rm E$, and their respective tangent linear mappings $\phi_0 : dx_0 \rightarrow dx$ from ${\rm T}_{x_0}(\rm S_{0,bf})$ to ${\rm T}_x(\rm S_{bf})$, $\phi : dx \rightarrow dx'$ from ${\rm T}_x(\rm S_{bf})$ to ${\rm T}_{x'}(\rm S'_{bf})$, $\iota_0 : dx_0 \rightarrow dx_0$ from ${\rm T}_{x_0}(\rm S_{0,bf})$ to $\rm E$, $\iota : dx \rightarrow dx$ from ${\rm T}_x(\rm S_{bf})$ to $\rm E$, $\psi_0 : dx_0 \rightarrow du$ from ${\rm T}_{x_0}(\rm S_{0,bf})$ to $\rm E$ and $\psi : dx \rightarrow dw$ from ${\rm T}_x(\rm S_{bf})$ to $\rm E$.

For arbitrary vectors $dx_0$ and $dy_0 \in {\rm T}_{x_0}(\rm S_{0,bf})$, the Lagrangian surface strain tensor is defined by
\begin{eqnarray}
&&e_{\rm s}(dx_0,dy_0) = \frac{1}{2} (dx \cdot dy - dx_0 \cdot dy_0) \nonumber\\
&=&\frac{1}{2} (dx^* \cdot dy - dx_0^* \cdot dy_0) \nonumber\\
&=&\frac{1}{2} (dx_0^* \cdot (\iota_0 + \psi_0)^* \cdot (\iota_0 + \psi_0) \cdot dy_0 
- dx_0^* \cdot \iota_0^* \cdot \iota_0 \cdot dy_0)
\end{eqnarray}
(see footnote\footnote{e.g., $dx^*$ is the linear form associated to $dx$, $\psi_0^* : {\rm E} \rightarrow {\rm T}_{x_0}(\rm S_{0,bf})$ is the adjoint of $\psi_0$.}),
which gives
\begin{eqnarray}
e_{\rm s} = \frac{1}{2} (\psi_0^* \cdot \iota_0 + \iota_0^* \cdot \psi_0 + \psi_0^* \cdot \psi_0)
\end{eqnarray}
($e_{\rm s}$ being here considered as an endomorphism of ${\rm T}_{x_0}(\rm S_{0,bf})$), i.e., using the coordinates
\begin{eqnarray}
e_{\rm s,\alpha \beta} = \frac{1}{2} (\partial_{0\alpha} u_i\,\partial_{0\beta} x_0^i
+ \partial_{0\beta} u_i\,\partial_{0\alpha} x_0^i
+ \partial_{0\alpha} u_i\,\partial_{0\beta} u^i)
\end{eqnarray}
(as a covariant tensor).
For any vectors $dx$ and $dy \in {\rm T}_x(\rm S_{bf})$, the Eulerian infinitesimal surface strain tensor $\delta \varepsilon_{\rm s}$ is defined by
\begin{eqnarray}
\delta \varepsilon_{\rm s} (dx,dy) &=& \frac{1}{2} \delta(dx \cdot dy) \nonumber\\
&=& \frac{1}{2} (\delta(dx) \cdot dy + dx \cdot \delta(dy)) \nonumber\\
&=& \frac{1}{2} (dx^* \cdot \psi^* \cdot \iota \cdot dy 
+ dx^* \cdot \iota^* \cdot \psi \cdot dy), 
\end{eqnarray}
which gives
\begin{eqnarray}
\delta \varepsilon_{\rm s} = \frac{1}{2} (\psi^* \cdot \iota + \iota^* \cdot \psi)\label{deltaepsilons}
\end{eqnarray}
($\delta \varepsilon_{\rm s}$ as an endomorphism of ${\rm T}_x(\rm S_{bf})$), i.e.,
\begin{eqnarray}
\delta \varepsilon_{\rm s,\alpha\beta} =
\frac{1}{2} (\partial_\alpha w_i\,\partial_\beta x^i
+ \partial_\beta w_i\,\partial_\alpha x^i)
\end{eqnarray}
(as a covariant tensor). It is related to the variation $\delta e_s$ of $e_s$ through
\begin{eqnarray*}
\delta e_{\rm s}(dx_0,dy_0) = \frac{1}{2} \delta(dx \cdot dy) = \delta \varepsilon_{\rm s} (dx,dy),
\end{eqnarray*}
i.e.,
\begin{eqnarray*}
dx_0^* \cdot \delta e_{\rm s} \cdot dy_0 &=& dx^* \cdot \delta \varepsilon_{\rm s} \cdot dy \\
&=& dx_0^* \cdot \phi_0^* \cdot \delta \varepsilon_{\rm s} \cdot \phi_0 \cdot dy_0,
\end{eqnarray*}
which gives
\begin{eqnarray}
\delta e_{\rm s} &=& \phi_0^* \cdot \delta \varepsilon_{\rm s} \cdot \phi_0 \nonumber\\
\delta \varepsilon_{\rm s} &=& \phi_0^{-1*} \cdot \delta e_{\rm s} \cdot \phi_0^{-1} \label{des}
\end{eqnarray}
($\delta e_{\rm s}$ and $\delta \varepsilon_{\rm s}$ as endomorphisms), i.e.,
\begin{eqnarray}
\delta e_{\rm s,\alpha\beta} &=&
\partial_{0\alpha} x^\zeta\,\partial_{0\beta} x^\eta\,\delta \varepsilon_{\rm s,\zeta\eta} \nonumber\\
\delta \varepsilon_{\rm s,\alpha\beta} &=&
\partial_\alpha x_0^\zeta\,\partial_\beta x_0^\eta\,\delta e_{\rm s,\zeta\eta}
\end{eqnarray}
(as covariant tensors).

Let $dx_0$ and $dy_0$ be arbitrary vectors of ${\rm T}_{x_0}(\rm S_{0,bf})$, $dl_0$ and $dl$ the respective lengths of $dx_0$ and $dx$, $\nu_0 \in {\rm T}_{x_0}(\rm S_{0,bf})$ a unit vector normal to
$dx_0$, and $\nu \in {\rm T}_x(\rm S_{bf})$ the unit vector normal to $dx$ such that $\nu \cdot (\phi_0 \cdot \nu_0) > 0$. The relation between the areas $da_0$ and $da$ of the two parallelograms respectively built with $(dx_0, dy_0)$ and $(dx, dy)$
\begin{eqnarray*}
A\,\nu_0\,dl_0 \cdot dy_0 = \nu\,dl \cdot dy,
\end{eqnarray*}
where $A = \displaystyle\frac{da}{da_0} = | \det (I \cdot \phi_0) |$ for any isometry $I : {\rm T}_x(\rm S_{bf}) \rightarrow {\rm T}_{x_0}(\rm S_{0,bf})$, gives
\begin{eqnarray}
\nu_0\,dl_0 = A^{-1}\,\phi_0^* \cdot \nu\,dl.\label{nudl}
\end{eqnarray}
We then define the Eulerian surface stress tensor $\sigma_{\rm s}$ such that the Eulerian surface stress force $\sigma_{\rm s} \cdot \nu\,dl$ results from the transport by $\phi_0$ of the Lagrangian surface stress force $\pi_{\rm s} \cdot \nu_0\,dl_0$:
\begin{eqnarray*}
\phi_0 \cdot \pi_{\rm s} \cdot \nu_0\,dl_0 = \sigma_{\rm s} \cdot \nu\,dl,
\end{eqnarray*}
which gives, according to (\ref{nudl}),
\begin{eqnarray}
\pi_{\rm s} &=& A\,\phi_0^{-1} \cdot \sigma_{\rm s} \cdot \phi_0^{-1*} \nonumber\\
\sigma_{\rm s} &=& A^{-1}\,\phi_0 \cdot \pi_{\rm s} \cdot \phi_0^* \label{pis}
\end{eqnarray}
($\pi_{\rm s}$ and $\sigma_{\rm s}$ as endomorphisms of ${\rm T}_{x_0}(\rm S_{0,bf})$ and ${\rm T}_x(\rm S_{bf})$, respectively), i.e.,
\begin{eqnarray}
\pi_{\rm s}^{\alpha\beta} &=&
A\,\partial_\zeta x_0^\alpha\,\partial_\eta x_0^\beta\,\sigma_{\rm s}^{\zeta\eta}\nonumber\\
\sigma_{\rm s}^{\alpha\beta} &=&
A^{-1}\,\partial_{0\zeta} x^\alpha\,\partial_{0\eta} x^\beta\,\pi_{\rm s}^{\zeta\eta}
\end{eqnarray}
(as contravariant tensors).

From (\ref{pis}) and (\ref{des}), we have
\begin{eqnarray*}
{\rm tr}(\pi_{\rm s} \cdot \delta e_{\rm s}) 
&=& A\,{\rm tr}(\phi_0^{-1} \cdot \sigma_{\rm s} \cdot \delta \varepsilon_{\rm s} \cdot \phi_0)\\
&=& A\,{\rm tr}(\sigma_{\rm s} \cdot \delta \varepsilon_{\rm s}),
\end{eqnarray*}
which leads to the Lagrangian and the Eulerian forms of the work of deformation of a surface element
\begin{eqnarray}
\pi_{\rm s} : \delta e_{\rm s}\,da_0 = \sigma_{\rm s} : \delta \varepsilon_{\rm s}\,da, \label{workdefsurf}
\end{eqnarray}
i.e.,
\begin{eqnarray*}
\pi_{\rm s}^{\alpha\beta}\,\delta e_{\rm s,\alpha\beta}\,da_0 
= \sigma_{\rm s}^{\alpha\beta}\,\delta \varepsilon_{\rm s,\alpha\beta}\,da.
\end{eqnarray*}

\section{Surface equations} \label{secSurfeq}

According to (\ref{deltaepsilons}), the last term of the equilibrium condition (\ref{Surfvar1}) is first written as 
\begin{eqnarray}
{\rm tr}(\sigma_{\rm s} \cdot \delta \varepsilon_{\rm s})
&=& {\rm tr}(\frac{\sigma_{\rm s}^* + \sigma_{\rm s}}{2} \cdot \iota^* \cdot \psi) \nonumber\\
&=& {\rm tr}(\sigma_{\rm s}^* \cdot \iota^* \cdot \psi)
+ {\rm tr}(\frac{\sigma_{\rm s} - \sigma_{\rm s}^*}{2} \cdot \iota^* \cdot \psi). \label{sigmas-sigmasstar}
\end{eqnarray}
$\rm S$ being a Riemannian manifold with boundary $\Gamma$, we then apply Green's formula \citep{Courrege:1966}
\begin{eqnarray*}
\int_{\rm S} {\rm div}X\,da = - \int_\Gamma X \cdot \nu\,dl
\end{eqnarray*}
(where $X$ is any vector field of class $C^1$ on $\rm S$---which is compact---, $\nu$ the field of unit vectors on $\Gamma$, tangent to $\rm S$, normal to $\Gamma$ and directed to the inside of $\rm S$, and $da$ and $dl$ are the Riemannian measures on $\rm S$ and $\Gamma$, respectively) to the vector field $X = \sigma_{\rm s}^* \cdot \iota^* \cdot w = {\widetilde{\sigma_{\rm s}}}^* \cdot w$, where $\widetilde{\sigma_{\rm s}} = \iota \cdot \sigma_{\rm s}$, if the components of $\sigma_{\rm s}$ and $w$ belong to $C^1(\rm S)$. At a point $x \in \rm S_{bf}$, ${\widetilde{\sigma_{\rm s}}}^* \in {\rm T}_x(\rm S_{bf}) \otimes E^*$ (linear mapping from $\rm E$ to ${\rm T}_x(\rm S_{bf})$), i.e., ${\widetilde{\sigma_{\rm s}}}^*$ is a section of the vector bundle $\rm T(S_{bf}) \otimes (S_{bf} \times E^*)$ over $\rm S_{bf}$. In order to decompose the term ${\rm div}X = {\rm div}({\widetilde{\sigma_{\rm s}}}^* \cdot w)$, we first need to define a covariant derivative and a divergence for ${\widetilde{\sigma_{\rm s}}}^*$.

In a general way, let us define the covariant derivative $\nabla$ on the vector bundle $({\rm T(S_{bf})^*})^{\otimes q} \otimes {\rm T(S_{bf})}^{\otimes p} \otimes ({\rm S_{bf} \times E^*})^{\otimes s} \otimes ({\rm S_{bf} \times E})^{\otimes r}$ over $\rm S_{bf}$ (for any $p$, $q$, $r$, $s \geq 0$), as the tensorial product of the covariant derivative $\nabla$ on $({\rm T(S_{bf})^*})^{\otimes q} \otimes {\rm T(S_{bf})}^{\otimes p}$ (for the Levi--Civita connection) and the usual derivative $d$ on $({\rm S_{bf} \times E^*})^{\otimes s} \otimes ({\rm S_{bf} \times E})^{\otimes r} = {\rm S_{bf}} \times (({\rm E^*})^{\otimes s} \otimes {\rm E}^{\otimes r})$, i.e., by
\begin{eqnarray}
\nabla_X (U \otimes V) = (\nabla_X U) \otimes V + U \otimes (d_X V),\label{tensorproductderiv}
\end{eqnarray}
for any sections $X$ of $\rm T(S_{bf})$, $U$ of $({\rm T(S_{bf})^*})^{\otimes q} \otimes {\rm T(S_{bf})}^{\otimes p}$, and $V$ of $({\rm S_{bf} \times E^*})^{\otimes s} \otimes ({\rm S_{bf} \times E})^{\otimes r}$ (this definition may be justified by using local frames of $({\rm T(S_{bf})^*})^{\otimes q} \otimes {\rm T(S_{bf})}^{\otimes p}$ and $({\rm S_{bf} \times E^*})^{\otimes s}$ $\otimes ({\rm S_{bf} \times E})^{\otimes r}$). For any section $W$ of $({\rm T(S_{bf})^*})^{\otimes q} \otimes {\rm T(S_{bf})}^{\otimes p} \otimes ({\rm S_{bf} \times E^*})^{\otimes s} \otimes ({\rm S_{bf} \times E})^{\otimes r}$, the covariant differential of $W$ is then defined as the linear mapping $\nabla W : X \rightarrow \nabla_X W$, so that $\nabla W$ is a section of $({\rm T(S_{bf})^*})^{\otimes (q + 1)} \otimes {\rm T(S_{bf})}^{\otimes p} \otimes ({\rm S_{bf} \times E^*})^{\otimes s} \otimes ({\rm S_{bf} \times E})^{\otimes r}$. As an example, for a section $W$ of $\rm T(S_{bf}) \otimes (S_{bf} \times E^*)$, $\nabla W$ is a section of $\rm T(S_{bf})^* \otimes T(S_{bf}) \otimes (S_{bf} \times E^*)$ and, by contraction of the covariant index relative to $\rm T(S_{bf})^*$ and the contravariant index relative to $\rm T(S_{bf})$, we thus define ${\rm div}\,W$, which is a section of $\rm S_{bf} \times E^*$. With respect to a local chart of $\rm S_{bf}$ (coordinates $(x^\alpha)$; Greek indices $\alpha$, $\beta$, $\gamma$,... belong to $\{1,2\}$), with the associated frames of $\rm T(S_{bf})$ and $\rm T(S_{bf})^*$, and to a basis of the vector space $\rm E$ (coordinates $(x^i)$; Latin indices $i$, $j$,... belong to $\{1,2,3\}$), with the associated dual basis of $\rm E^*$, we may thus write the components
\begin{eqnarray}
(\nabla W)^\alpha_{\beta i} &=& \partial_\beta W^\alpha_i + \Gamma^\alpha_{\beta \gamma} W^\gamma_i\\
({\rm div}\,W)_i &=& \partial_\beta W^\beta_i + \Gamma^\beta_{\beta \gamma} W^\gamma_i, \label{divergencecomp}
\end{eqnarray}
where $\Gamma^\alpha_{\beta \gamma}$ are the Christoffel's symbols of the Levi--Civita connection on $\rm S_{bf}$.

With this definition, we may now write
\begin{eqnarray*}
\nabla ({\widetilde{\sigma_{\rm s}}}^* \otimes w) 
= (\nabla {\widetilde{\sigma_{\rm s}}}^*) \otimes w + {\widetilde{\sigma_{\rm s}}}^* \otimes \psi
\end{eqnarray*}
($\psi$ being the usual derivative of $w$), then
\begin{eqnarray*}
\nabla ({\widetilde{\sigma_{\rm s}}}^* \cdot w) = (\nabla {\widetilde{\sigma_{\rm s}}}^*) \cdot w 
+ {\widetilde{\sigma_{\rm s}}}^* \cdot \psi
\end{eqnarray*}
(contraction of the covariant index deriving from ${\widetilde{\sigma_{\rm s}}}^*$ and the contravariant index deriving from $w$) and
\begin{eqnarray}
{\rm div} ({\widetilde{\sigma_{\rm s}}}^* \cdot w) = {\rm div} ({\widetilde{\sigma_{\rm s}}}^*) \cdot w 
+ {\widetilde{\sigma_{\rm s}}}^* : \psi
\end{eqnarray}
(contraction of the covariant index deriving from $\nabla$ and the contravariant index deriving from ${\widetilde{\sigma_{\rm s}}}^*$). Green's formula applied to $X = {\widetilde{\sigma_{\rm s}}}^* \cdot w$ may then be written as
\begin{eqnarray}
\int_{\rm S} {\rm tr}(\sigma_{\rm s}^* \cdot \iota^* \cdot \psi)\,da
&=& - \int_{\rm S} {\rm div}({\widetilde{\sigma_{\rm s}}}^*) \cdot w\,da
- \int_{\Gamma} ({\widetilde{\sigma_{\rm s}}}^* \cdot w) \cdot \nu\,dl \nonumber\\
&=& - \int_{\rm S} {\rm div}({\widetilde{\sigma_{\rm s}}}^*) \cdot w\,da 
- \int_{\Gamma} (\sigma_{\rm s} \cdot \nu) \cdot w\,dl. \label{Greensurface} 
\end{eqnarray}
The condition (\ref{Surfvar2}) is then obtained, from (\ref{sigmas-sigmasstar}) and (\ref{Greensurface}) (since $w = 0$ on $\Gamma$).

Let us now consider $\sigma_{\rm s}$ as a contravariant tensor (convention used in (\ref{bfequilibrium1comp})), denote $\bar{\sigma_{\rm s}} = \iota \cdot \sigma_{\rm s}$ (by contraction of the covariant index of $\iota$ and the first contravariant index of $\sigma_{\rm s}$) and write
\begin{eqnarray*}
\nabla (\iota \otimes \sigma_{\rm s}) = (\nabla \iota) \otimes \sigma_{\rm s} + \iota \otimes (\nabla \sigma_{\rm s}),
\end{eqnarray*}
hence
\begin{eqnarray*}
\nabla \bar{\sigma_{\rm s}} = (\nabla \iota) \cdot \sigma_{\rm s} + \iota \cdot (\nabla \sigma_{\rm s})
\end{eqnarray*}
(contraction of the covariant index deriving from $\iota$ and the first contravariant index deriving from $\sigma_{\rm s}$) and
\begin{eqnarray}
{\rm div}\,\bar{\sigma_{\rm s}} = l : \sigma_{\rm s} + \iota \cdot {\rm div}\,\sigma_{\rm s} \label{divdecompos}
\end{eqnarray}
(contraction of the covariant index deriving from $\nabla$ and the second contravariant index deriving from $\sigma_{\rm s}$; ${\rm div}\,\sigma_{\rm s}$ is the usual surface divergence), where
\begin{eqnarray}
l = \nabla \iota, 
\end{eqnarray}
i.e., with the components,
\begin{eqnarray}
({\rm div}\,\bar{\sigma_{\rm s}})^i = \sigma_{\rm s}^{\alpha\beta}\,l_{\alpha\beta}^i + ({\rm div}\,\sigma_{\rm s})^\alpha\,\partial_\alpha x^i, \label{divdecomposcomp}
\end{eqnarray}
with
\begin{eqnarray}
({\rm div}\,\bar{\sigma_{\rm s}})^i &=& \partial_\beta (\sigma_{\rm s}^{\alpha\beta}\,\partial_\alpha x^i) +
\Gamma_{\beta\gamma}^\beta\,\sigma_{\rm s}^{\alpha\gamma}\,\partial_\alpha x^i\nonumber\\
l_{\alpha\beta}^i &=& \partial_{\alpha\beta} x^i - \Gamma_{\alpha\beta}^\gamma\,\partial_\gamma x^i\nonumber\\
({\rm div}\,\sigma_{\rm s})^\alpha\ &=& \partial_\beta \sigma_{\rm s}^{\alpha\beta} + 
\Gamma_{\beta\gamma}^\alpha\,\sigma_{\rm s}^{\gamma\beta} + 
\Gamma_{\beta\gamma}^\beta\,\sigma_{\rm s}^{\alpha\gamma}.
\end{eqnarray}
Moreover, for any sections $X$ and $Y$ of $\rm T(S_{bf})$, we have
\begin{eqnarray*}
\nabla_X (\iota \otimes Y) = (\nabla_X \iota) \otimes Y + \iota \otimes (\nabla_X Y),
\end{eqnarray*}
hence\begin{eqnarray*}
\nabla_X (\iota \cdot Y) = (\nabla_X \iota) \cdot Y + \iota \cdot (\nabla_X Y)
\end{eqnarray*}
(contraction of the covariant index deriving from $\iota$ and the contravariant index deriving from Y), i.e.
\begin{eqnarray}
d_X Y &=& (\nabla_X \iota) \cdot Y + \nabla_X Y \nonumber\\
&=& l : (X \otimes Y) + \nabla_X Y\label{secondform}
\end{eqnarray}
($Y$ and $\nabla_X Y$ being identified to $\iota \cdot Y$ and $\iota \cdot (\nabla_X Y)$, respectively), which shows that $l$ is the second vectorial fundamental form on $\rm S_{bf}$ (see \cite{Dieudonne:1971}, (20.12.4)), so that (\ref{divdecompos}) and (\ref{secondform}) respectively represent the decomposition of ${\rm div}\,\bar{\sigma_{\rm s}}$ and $d_X Y$ into the normal component ($l : \sigma_{\rm s}$ and $l : (X \otimes Y)$, respectively) and the tangential component (${\rm div}\,\sigma_{\rm s}$ and $\nabla_X Y$, respectively), with respect to $\rm S_{bf}$. This leads to (\ref{bfequilibrium1tangent}) and (\ref{bfequilibrium1normal}).

\section*{Dedication}
Dedicated to the memory of my mother.


\end{document}